\documentclass[12pt,draftclsnofoot,letterpaper,onecolumn,romanappendices]{IEEEtran}
 \usepackage{cite}
 \usepackage{epstopdf}
 \usepackage{epsfig}
 \usepackage{amsthm}
 \usepackage{amsmath,amssymb,amsfonts,amstext,amsbsy,amsopn,dsfont}
 \usepackage{cases}
 \usepackage{bigints}
 \usepackage{sublabel}
 \usepackage{color}
 \usepackage{array}
 
\newtheorem{proposition}{\textbf{Proposition}}

\newtheorem{remark}{\textbf{Remark}}

\newcommand{\defn}{\triangleq}
\newcommand{\dif}{\textmd{d}}
\newcommand{\ie}{\emph{i.e.,} }

\begin{document}

\title{\LARGE Heterogeneous Networks with Power-Domain NOMA: Coverage, Throughput and Power Allocation Analysis}

\author{Chun-Hung Liu and Di-Chun Liang
\thanks{C.-H. Liu is with the Department of Electrical and Computer Engineering, University of Michigan, MI 48128, USA. D.-C. Liang is with the Institute of Communications Engineering and Department of Electrical and Computer Engineering, National Chiao Tung University, Hsinchu 30010, Taiwan. Part of this work was presented at the IEEE Vehicular Technology Conference, Spring 2017 \cite{CHLDCLPCCJRY17}. Dr. Liu is the contact author of this paper (e-mail: chunhunl@umich.edu). }}
\maketitle

\begin{abstract}
In a heterogeneous cellular network (HetNet), consider that a base station in the HetNet is able to simultaneously schedule and serve $K$ users in the downlink by performing the power-domain non-orthogonal multiple access (NOMA) scheme. This paper aims at the preliminary study on the downlink coverage and throughput performances of the HetNet with the non-coordinated NOMA and the proposed coordinated joint transmission NOMA (JT-NOMA) schemes. First, the coverage probability and link throughput of $K$ users in each cell are studied and their accurate expressions are derived for the non-coordinated NOMA scheme in which no BSs are coordinated to jointly transmit the NOMA signals for a particular user. We show that the coverage and link throughput can be largely reduced if transmit power allocations among the $K$ users do not satisfy the constraint derived. Next, we analyze the coverage and link throughput of $K$ users for the coordinated JT-NOMA scheme in which the void BSs without users are coordinated to enhance the farthest NOMA user in a cell. The derived accurate results show that coordinated JT-NOMA can significantly improve the coverage and link throughput of all users. Finally, we show that there exist optimal power allocation schemes that maximize the average cell coverage and throughput under some derived power allocation constraints and numerical results validate our analytical findings.   
\end{abstract}

\section{Introduction}
In traditional cellular networks, orthogonal multiple access (OMA) schemes, such as frequency
division multiple access (FDMA), time division multiple access (TDMA) and code division multiple access (CDMA), are able to successfully suppress a large amount of co-channel interferences so that the signal-to-interference power ratio  (SIR) on the receiver side can be enhanced remarkably. However, enhancing SIR via OMA is not the most efficient/effective method to improve the spectrum efficiency of a wireless link in an interference-limited network according to the fundamental result of the multiuser capacity region \cite{DTPV05}\cite{LDBWYYSHCLIZW15}. To meet the huge throughput need anticipated in 5G cellular networks under the pressing pressure of spectrum crunch, non-orthogonal multiple access (NOMA)  has gained a lot of attentions recently \cite{ABYSYKALAHTN13,LDBWYYSHCLIZW15,SMRNAOADKSK17,ZDYLJCQSME17} in that it is able to make the scarce spectrum resource be utilized and shared in a more efficient fashion as well as  reduce the complexity in resource allocation and user scheduling.  

It is well known that in a downlink cellular network the power-multiplexing NOMA scheme\footnote{The NOMA scheme in this paper is a multiplexing superposition coding scheme in the power domain \cite{LDBWYYSHCLIZW15,SMRNAOADKSK17,ZDYLJCQSME17}, \ie different downlink users are allocated different powers based on their channel gain conditions, whereas other code-domain-multiplexing NOMA schemes are beyond the scope of this work.} that adopts successive interference cancellation  (SIC) to perfectly cancel the multiuser interference always achieves a larger sum throughput (achievable rate) than the OMA schemes. However, the SIR performance of each individual (NOMA) user is definitely degraded due to power sharing among multiple users. In a heterogeneous cellular network (HetNet), the coverage (probability), \ie the probability that the SIR of users in the network is higher than some predesignated threshold, actually dominates the performances of all SIR-related metrics so that the link capacity of each user cannot be improved provided users' coverage is severely degraded. The coverage-degraded problem for the users turns out to be even much worse in an interference-limited HetNet using NOMA where a large amount of interference is generated by many different kinds of densely-deployed base stations (BSs) and users have to share the transmit power of a BS. Accordingly, how the SIR of users behaves in a HetNet using NOMA is an important topic that needs to be studied thoroughly.

\subsection{Motivation and Related Prior Work}
The prior works on the study of the coverage/outage, link throughput and power allocation  problems in a large-scale NOMA HetNet with multicell interferences are still minimal. Few current works have studied the transmission performance of the NOMA scheme in a cellular network based on a single cell model. For example, reference \cite{ZDZYPFHVP14} studied the performance of the outage and ergodic rate of the NOMA scheme in a single cell and showed that NOMA can achieve a higher sum rate whereas its rate gain in the low SNR region is not significant. In \cite{YLZDMEVP16}, a cooperative NOMA scheme was proposed to simultaneously transfer wireless information and power for users in a single cell and the outage probabilities for different user selection schemes were analyzed. The idea of achieving cooperative NOMA transmission is to let the users that are close to their BS and have good channel conditions relay the weak signals of the users that are fairly far away from their BS. Some other prior works, such as \cite{ZDMPHVP15,ZYZDPFNAD17,DWMWFJYLYH18}, also adopted the similar idea of exploiting the collaboration between users to achieve cooperative NOMA. Reference  \cite{YZHMWTXZ17} looked into the energy-efficient power allocation problem for NOMA and reference \cite{ZDZYPFHVP17} studied how to enhance the spectrum efficiency and security in a multiuser network with mixed multicasting and unicasting traffic. 

These aforementioned works are not studied in a large-scale multicell HetNet and thus generally they are unable to project how the coverage and throughput of users are impacted by multicell interferences. There are few prior NOMA works that are developed based on a large-scale multicell network model. In \cite{YLYKMQYYHLZ17}, for example, the outage probability was studied in a large-scale cognitive radio network. Similarly, reference \cite{ZZHSRQH17} analyzed the outage and achievable rate of users in a single-tier cellular network. In \cite{YLZQMEAN17}, the NOMA scheme was considered to be performed in a HeNet, and then the coverage, ergodic rate and energy efficiency performances were analyzed. The cooperative multicast problem in a NOMA mmWave HetNet was considered in \cite{ZZZMYX17} and the multicast rate was shown to be significantly improved by NOMA in this work. These prior works do not address the problem of how power allocations among the users affect the outage and rate performances in a large-scale network environment. Accordingly, while using power-domain NOMA in a HetNet, how to appropriately allocate different powers to different users in order to improve the coverage and link throughput performances of the users are still not very much clear.   

\subsection{Contributions}
To investigate the fundamental interactions among coverage, link throughput and power allocation of NOMA transmissions in a large-scale HetNet, in this paper we consider a HetNet in which BSs that are associated with multiple users can perform the NOMA scheme to serve their scheduled users. Our first contribution is to construct a stochastic-geometry-based HetNet model in which users associate with their BS using a biased nearest BS association scheme and each BS with multiple tagged users is able to schedule  at most $K$ users for downlink NOMA transmission. We propose the ``desired" SIR model for the $K$-user NOMA scheme. This model considers the impact of the void BSs that are not associated with any users so that it is more accurate especially when the HetNet is dense (\ie the user intensity is not significantly smaller than the total BS intensity.) \cite{CHLLCW15,CHLLCW16}. Under this HetNet model, we first consider that each BS  which can arbitrarily schedule $K$ users adopts the non-coordinated NOMA scheme in which no void BSs in the network are coordinated to help a non-void BS jointly transmit its NOMA signals. For this non-coordinated NOMA scheme, we derive the explicit expressions of the coverage probability and link throughput of a user associating with a tier-$m$ BS by successfully characterizing the channel ordering statistics of the scheduled NOMA users. They are in general very accurate and approach to their theoretical exact expressions as the user intensity goes to infinity. According to the derived results, we characterize some fundamental power allocation constraints for successfully performing the non-coordinated NOMA scheme and facilitating the derivations of the analytical results. We also show that the coverage and link throughput performances are dominated by power allocations among the $K$ users and the sum of the link throughputs of all NOMA users is strictly higher than the link throughput of a sole user that uses the full transmit power of its tagged BS. This is our second contribution. 

To alleviate the impact of the power allocations among all NOMA users, our third contribution is to propose the coordinated JT-NOMA scheme in which all void BSs are coordinated to do joint transmission of the farthest user in a particular cell. This coordinated JT-NOMA scheme not only helps the near users do SIC but also helps the farthest user decode its own signals. Therefore, the SIR performances of the NOMA users are all improved. Note that the coordinated JT-NOMA scheme which is a BS-level cooperative scheme is essentially different from the cooperative NOMA schemes in the literature that are a user-level cooperative scheme \cite{ZDMPHVP15,YLZDMEVP16,ZYZDPFNAD17,DWMWFJYLYH18}. The accurate expressions of the coverage and link throughput of each user associating with a tier-$m$ BS are found. They clearly indicate how the coordinated JT-NOMA scheme achieves higher coverage as well as link throughput and how 
power allocations among the users influence the coverage and link throughput. Most importantly, they characterize some fundamental power allocation constraints that make coordinated JT-NOMA perform well and facilitate the analyses.

Afterwards we analyze how to optimally allocate the powers among the $K$ NOMA users for the non-coordinated NOMA and coordinated JT-NOMA schemes in order to maximize the cell coverage and the cell throughput of each BS. The optimization problems of the tier-$m$ cell coverage and cell throughput are formulated based on the power allocation constraints found while analyzing coverage and link throughput. We show that an optimal power allocation scheme indeed exists for the two formulated optimization problems that are not convex in general and it can be found by some heuristic optimization algorithms. Note that the power allocation problems studied are network-based optimization problems since they are formulated based on a large-scale multicell HetNet, which are different from the single-cell-based power allocation problems in almost all the prior works (typically see \cite{ZDXLGKK17,PXKC17,STIK15}). This summarizes our fourth contribution. Finally, some numerical results are provided to validate our analytical findings and observations.     

\subsection{Paper Organization}
The rest of this paper is organized as follows. Section \ref{Sec:SystemModel} introduces a multi-tier heterogeneous network model as well as some important assumptions. In Section \ref{Sec:NonCoopNOMA}, the downlink coverage and link throughput are analyzed for the scenario that no void BSs are coordinated to do joint NOMA transmission. For the scenario that all void BSs are coordinated to do joint NOMA transmission, the downlink coverage and link throughput are studied in Section \ref{Sec:CoopNOMA}. The optimal power allocation problem is formulated and investigated in Section \ref{Sec:OptimalPowerAllocation}. In Section \ref{Sec:Simulation}, some numerical results are provided to validate our analytical findings in coverage, link throughput and power allocation. Finally, Section \ref{Sec:Conclusions} briefly summarizes our analytical achievements and observations for a HetNet using NOMA transmission.

\section{Network Model and Assumptions}\label{Sec:SystemModel}
Consider a large-scale interference-limited HetNet on the plane $\mathbb{R}^2$ in which there are $M$ different types of base stations BSs (e.g., macrocell, microcell, picocell BSs, etc.) and the BSs of each type are referred as a tier of the HetNet. Specifically, we assume that the BSs in the $m$th tier form an independent homogeneous Poisson point process (PPP) of intensity $\lambda_m$ and they are denoted by set $\Phi_m$ that can be explicitly written as
\begin{align}
\Phi_m\defn \{X_{m,i}\in\mathbb{R}^2 : i\in\mathbb{N}_+\}, \,\,m\in\mathcal{M}\defn\{1,2,\ldots,M\},
\end{align}
where $X_{m,i}$ denotes BS $i$ in the $m$th tier and its location. All users are also assumed to form an independent PPP of intensity $\mu$ and they are denoted by set $\mathcal{U}\subseteq \mathbb{R}^2$. Let $\mathcal{U}_{m,i}$ denote the set of the users associating with BS $X_{m,i}$ and it can be expressed as
\begin{align}\label{Eqn:UserSetX_m}
\mathcal{U}_{m,i}\defn\left\{U_n\in\mathcal{U}: X_{m,i}=\arg\sup_{l,j:X_{l,j}\in\Phi}\left\{\omega_l\|X_{l,j}-U_n\|^{-\alpha}\right\}, n, j\in\mathbb{N}_+,l\in\mathcal{M}\right\},
\end{align}
where $\Phi\defn\bigcup_{m=1}^M\Phi_m$ is the set of all BSs in the HetNet, $\alpha>2$ is the pathloss exponent, $\omega_l>0$ is the (constant) user association bias for the tier-$l$ BSs\footnote{The function of the user association bias ($\omega_l$) for every tier is used for traffic offloading/loading or cell range expansion in order to make the cell load of each BS achieve a certain level of balancing.}. Note that the biased nearest BS association (BNBA) scheme is adopted in \eqref{Eqn:UserSetX_m}, \ie this BNBA scheme makes users select their nearest BS with a particular bias for each tier\footnote{To make the following analysis much more tractable, in this paper a constant bias is used for each tier so that the BSs in each tier have a weighted Voronoi-tessellated cell. More general user association schemes with a random bias for each tier, such as maximum received-power association and energy-efficient user association, can be referred to our previous works in \cite{CHLLCW16,CHLKLF16}.}.  

\subsection{User Association and Downlink NOMA Transmission}
Let $|\mathcal{U}_{m,i}|$ denote the cardinality of set $\mathcal{U}_{m,i}$, \ie the number of the users associating with  BS $X_{m,i}$. The probability mass function (pmf) of $|\mathcal{U}_{m,i}|$, based on our previous work in \cite{CHLKLF16}, is approximately found as
\begin{align}\label{Eqn:pmfNumUsers}
p_{m,n}\defn\mathbb{P}[|\mathcal{U}_{m,i}|=n]\approx\frac{\Gamma(n+\frac{7}{2})}{n! \Gamma(\frac{7}{2})}\left(\frac{2}{7}\xi_m\right)^{n}\left(1+\frac{2}{7}\xi_m\right)^{-(n+\frac{7}{2})},
\end{align}   
where $\xi_m\defn \frac{\mu\omega^{\frac{2}{\alpha}}_m}{\sum_{l=1}^{M}\omega^{\frac{2}{\alpha}}_l\lambda_l}$ is called the \textit{cell load} of a tier-$m$ BS for the BNBA scheme and it represents the mean number of users associating with a tier-$m$ BS. According to \eqref{Eqn:pmfNumUsers}, we know $p_{m,0}\approx (1+\frac{2}{7}\xi_m)^{-\frac{7}{2}}$, which is called the \textit{tier-$m$ void (cell) probability}, \ie the probability that a tier-$m$ BS is not associated with any users. In other words, the non-void probability of a tier-$m$ BS in the HetNet can be readily written as
\begin{align}\label{Eqn:NonVoidProbTier-m}
\nu_m \defn 1-p_{m,0} \approx1-\left(1+\frac{2}{7}\xi_m\right)^{-\frac{7}{2}}.
\end{align}
Note that the non-void probability $\nu_m$ is small as the user intensity is not much smaller than the total intensity of BSs, for example, the intensity of a dense HetNet is close to or even larger than the user intensity\footnote{In this paper, our study will focus on the scenario of a ``dense'' HetNet in which small cell BSs, such as picocell and femtocell BSs, are deployed with a high intensity and their intensities may not be smaller than the user intensity.}. As a result, the intensity of the void BSs, $\sum_{m=1}^{M}p_{m,0}\lambda_m$, is not small in a dense HetNet, which means the void cell phenomenon that is usually overlooked in the literature should be carefully considered in the interference model of a dense HetNet in that those void BSs are actually idle and do not generate interference \cite{CHLLCW15,CHLLCW16}. 

For the BSs having at least two users, they can use the NOMA scheme with \textit{superposition coding} to simultaneously transmit different data streams to different users over the same frequency band. The NOMA scheme considered in this paper is performed in the power domain, \ie the BSs allocate different transmit powers to different users while transmitting according to the channel conditions of their users \cite{LDBWYYSHCLIZW15}, and then users are able to perform SIC to decode their own data. To tractably and simply analyze the downlink SIR of a user, we specifically consider that \textit{each BS is able to arbitrarily schedule at most $K$ NOMA users even if it is associated with more than $K$ users}. For a BS only having a single user, it just transmits data to its sole user with full transmit power. In this paper, we will study two NOMA schemes: \textit{coordinated JT-NOMA and non-coordinated NOMA schemes}. For the coordinated JT-NOMA scheme, we assume that the void BSs can be coordinated to help other non-void BSs to simultaneously transmit the signals of the non-void BSs to their NOMA users, whereas for the non-coordinated NOMA scheme no void BSs are coordinated to do joint transmission. Hence, coordinated JT-NOMA is essentially a scheme of joint-transmission coordinated multipoint (JT-CoMP) \cite{PXCHLJGA13}. The non-coordinated NOMA and coordinated JT-NOMA schemes will be investigated and discussed in Sections \ref{Sec:NonCoopNOMA} and \ref{Sec:CoopNOMA}, respectively.

\subsection{The Desired SIR Model for Downlink NOMA transmission}

Suppose BS $X_{m,i}$ is associated with at least $K$ users so that it is able to schedule $K$ users for downlink NOMA transmission in each time slot. Without loss of generality, consider BS $X_{m,i}$ located at the origin\footnote{According to the Slivnyak theorem \cite{DSWKJM13,MH12,FBBBL10}, the statistical properties evaluated at any particular point in homogeneous PPPs are the same.} and let $U_k\in\mathcal{U}_{m,i}$ be the $k$th nearest user among the $K$ users scheduled by BS $X_{m,i}$. The ``desired" SIR at $U_k$ can be written as\footnote{This desired SIR of user $U_k$ is the SIR of user $U_k$ without considering the interferences from other NOMA users.}
\begin{align}\label{Eqn:DSIR}
\gamma_{m,k}\defn \frac{\beta_kP_mH_{m,i,k}}{\|U_k\|^{\alpha}I_{m,k}},\,k\in\mathcal{K}\defn\{1,2,\ldots,K\},
\end{align}
where $P_m$ is the total transmit power of a tier-$m$ BS, $\beta_kP_m$ is the transmit power allocated to user $U_k$, $\beta_k\in(0,1)$ is the power allocation fraction for $U_k$ associating with a tier-$m$ BS and thus $\sum_{k=1}^{K}\beta_k=1$, $H_{m,i,k}$ is the Rayleigh fading channel gain from BS $X_{m,i}$ to user $U_k$, $\|Y'_i-Y'_j\|$ denotes the Euclidean distance between nodes $Y'_i$ and $Y'_j$, $I_{m,k}$ is the interference received by $U_k$ and it is given by
\begin{align*}
I_{m,k}\defn \sum_{l,j:X_{l,j}\in\Phi\setminus X_{m,i}} V_{l,j}P_lH_{l,j,k} \|X_{l,j}-U_k\|^{-\alpha}
\end{align*}  
in which $V_{l,j}\in\{0,1\}$ is a Bernoulli random variable that is one if BS $X_{l,j}$ is non-void and zero otherwise. Throughout this paper, all fading channel gains are assumed to be i.i.d. exponential random variables with unit mean and variance, \ie $H_{m,i,k}\sim\mathrm{Exp}(1)$ for all $m\in\mathcal{M}$, $i\in\mathbb{N}_+$ and $k\in\mathcal{K}$, and the shadowing effect on all channels is ignored to facilitate the following analysis. Moreover, since $\|U_k\|\leq\|U_{k+1}\|$ for all $k\in\mathcal{K}$,  we have to let power allocation fractions  of the $K$ users follow by the constraint $\beta_1<\cdots<\beta_k< \cdots< \beta_K$ such that the farther users are allocated more transmit power. Such a power allocation constraint not only facilitates the SIC performed by the nearer users but also characterizes the resource allocation fairness among users.

The complementary cumulative density function (CCDF) of $\gamma_{m,k}$ in \eqref{Eqn:DSIR} has a tight and explicit lower bound, as shown in the following proposition.
\begin{proposition}\label{Prop:CCDF-DSIR-Tier-m}
For a given $x>0$, the CCDF of the desired SIR of user $U_k\in\mathcal{U}_{m,i}$ in \eqref{Eqn:DSIR}, i.e., $F^{\mathsf{c}}_{\gamma_{m,k}}(x)\defn\mathbb{P}\left[\gamma_{m,k}\geq x\right]$, has a tight lower bound given by
\begin{align}\label{Eqn:CCDF-DSIR-mk}
F^{\mathsf{c}}_{\gamma_{m,k}}(x)\gtrapprox \prod_{j=0}^{k-1}\frac{(K-j)}{(K-j)+\sum_{l=1}^{M}\nu_l\ell_{m,l}(x/\beta_k)},\,\,k\in\mathcal{K},
\end{align} 
where $x \gtrapprox y$ denotes that $y$ is a tight lower bound on $x$ and $\ell_{m,l}(\cdot)$ is defined as
\begin{align}\label{Eqn:ellfun-mk}
\ell_{m,l}(x)\defn \varphi_l\left(\frac{x \omega_mP_l}{\omega_lP_m}\right)^{\frac{2}{\alpha}}\left(\frac{1}{\mathrm{sinc}(2/\alpha)}-\int_{0}^{\left(\frac{\omega_lP_m}{x\omega_m P_l}\right)^{\frac{2}{\alpha}}}\frac{\dif t}{1+t^{\frac{\alpha}{2}}}\right)
\end{align}
in which $\varphi_l\defn \omega^{\frac{2}{\alpha}}_l\lambda_l/\sum_{m=1}^{M}\omega^{\frac{2}{\alpha}}_m\lambda_m$ represents the probability that a user associates with a tier-$l$ BS by using the BNBA scheme and $\mathrm{sinc}(x)\defn\frac{\sin(\pi x)}{\pi x}$ is the (normalized) sinc function. 
\end{proposition}
\begin{IEEEproof}
See Appendix \ref{App:ProofPropCCDF-DSIR-Tier-m}.
\end{IEEEproof}

In general, the lower bound in \eqref{Eqn:CCDF-DSIR-mk} is very tight since it is derived by using the fact that the location correlations among all non-void BSs are in general pretty low, as pointed out in the proof of Proposition \ref{Prop:CCDF-DSIR-Tier-m}. As the user intensity becomes larger and larger (\ie the location correlations among the non-void BSs become weaker and weaker), $F^{\mathsf{c}}_{\gamma_{m,k}}(x)$ gradually and eventually reduces to the lower bound given by 
\begin{align}\label{Eqn:LowestBoundCCDFSIRmk}
\lim_{\mu\rightarrow\infty} F^{\mathsf{c}}_{\gamma_{m,k}}(x)=\prod_{j=0}^{k-1}\frac{(K-j)}{[(K-j)+\sum_{l=1}^{M}\ell_{m,l}(x/\beta_k)]}
\end{align}
because no void BSs exist in the HetNet (\ie $\nu_l=1$ for all $l\in\mathcal{M}$). This indicates that \eqref{Eqn:LowestBoundCCDFSIRmk} is the lowest limit on $F^{\mathsf{c}}_{\gamma_{m,k}}(x)$. Also,  $F^{\mathsf{c}}_{\gamma_{m,k}}(x)$ in \eqref{Eqn:CCDF-DSIR-mk} is valid for the CCDF results of other specific BNBA schemes since it is derived based on the BNBA scheme with general biases. In the unbiased NBA scheme, for instance, all $\omega_m$'s are the same (and they can be set as one) and thus \eqref{Eqn:ellfun-mk} becomes
\begin{align}\label{Eqn:ellfun-mkNBA}
\ell_{m,l}(x)=\varphi_l\left(\frac{x P_l}{P_m}\right)^{\frac{2}{\alpha}}\left(\frac{1}{\mathrm{sinc}(2/\alpha)}-\int_{0}^{\left(\frac{P_m}{x P_l}\right)^{\frac{\alpha}{2}}}\frac{\dif t}{1+t^{\frac{\alpha}{2}}}\right)
\end{align}
and $\varphi_l=\lambda_l/\sum_{m=1}^{M}\lambda_m$. Substituting \eqref{Eqn:ellfun-mkNBA} into \eqref{Eqn:CCDF-DSIR-mk} yields the CCDF of $\gamma_{m,k}$ for the unbiased NBA scheme. Another example is to designate $\omega_m=P_m$ for all $m\in\mathcal{M}$ and the BNBA scheme is essentially to make users associate with the BS that offers the average maximum received power to them. Such a scheme is called the maximum received power association (MRPA) and \eqref{Eqn:ellfun-mk} for this scheme becomes
\begin{align*}
\ell_{m,l}(x)=\varphi_lx^{\frac{2}{\alpha}}\left(\frac{1}{\mathrm{sinc}(2/\alpha)}-\int_{0}^{x^{-\frac{\alpha}{2}}}\frac{\dif t}{1+t^{\frac{\alpha}{2}}}\right),
\end{align*}
where $\varphi_l=P^{\frac{2}{\alpha}}_l\lambda_l/\sum_{m=1}^{M}P^{\frac{2}{\alpha}}_m\lambda_m$. In addition, the result in \eqref{Eqn:CCDF-DSIR-mk} reveals a pivotal phenomenon: the farther user, the lower CCDF of its desired SIR even though the farther users are allocated more powers. As we will show in the following analyses, this phenomenon dominates the properties of the coverage and link throughput of each NOMA user.

\section{Downlink Coverage and Throughput Analysis for non-coordinated NOMA}\label{Sec:NonCoopNOMA}
In this section, we would like to study the downlink coverage and link throughput of a user associating with a tier-$m$ BS that adopts the \textit{non-coordinated} NOMA scheme to simultaneously transmit multiple data streams to its multiple users. Understanding the coverage performance of each (NOMA) user is quiet important in that each user is only allocated with some fraction of the total transmit power and thus the coverage performance of each user must degrade under the NOMA scheme if compared with the coverage performance of single user (OMA) transmission. The coverage analysis provides us with some insight into how many users should be served by NOMA provided there is a coverage constraint that needs to be satisfied. Likewise, the link throughput performance of a user is also very crucial because it reflects how much (sum) throughput a BS using NOMA could achieve and how many users should be appropriately scheduled at the same time so as to efficiently improve the sum throughput of a BS. We start with the coverage analysis in the following subsection and then the link throughput analysis afterwards.

\subsection{Coverage Analysis for non-coordinated NOMA}\label{SubSec:CovAnaNonCooPNOMA}
Suppose BS $X_{m,i}$ is able to  schedule $K$ users in its tagged user set $\mathcal{U}_{m,i}$ for non-coordinated NOMA transmission. The coverage (probability) $\rho_{m,k} $ of the $k$th nearest scheduled user to BS $X_{m,i}$ is defined as
\begin{align}
\rho_{m,k}&\defn\mathbb{P}\left[\frac{\beta_kP_mH_{m,i,k}\|U_k\|^{-\alpha}}{(\sum_{n=0}^{k-1}\beta_n)P_m\frac{H_{m,i,k}}{\|U_k\|^{\alpha}}+I_{m,k}}\geq\theta,\cdots,\frac{\beta_KP_mH_{m,i,k}\|U_k\|^{-\alpha}}{(\sum_{n=0}^{K-1}\beta_n)P_m\frac{H_{m,i,k}}{\|U_k\|^{\alpha}}+I_{m,k}}\geq\theta\right] \nonumber\\ &=\mathbb{P}\left[\frac{\beta_k\gamma_{m,k}}{(\sum_{n=0}^{k-1}\beta_n)\gamma_{m,k}+\beta_k}\geq\theta,\cdots,\frac{\beta_K\gamma_{m,k}}{(\sum_{n=0}^{K-1}\beta_n)\gamma_{m,k}+\beta_k}\geq\theta\right],\label{Eqn:DefnCovProbUk} 
\end{align}
where $\theta>0$ is the SIR threshold for successful decoding and $\beta_{0}\defn0$. The definition of $\rho_{m,k}$ has to include the event of successfully decoding the signals of the $K-k$ users farther than $U_k$ by using SIC before $U_k$ can successfully decode its own signal.  The explicit result of $\rho_{m,k}$ is found in the following proposition.
\begin{proposition}\label{Prop:CovProbNOMA}
If a tier-$m$ BS is able to arbitrarily schedule $K$ NOMA users and the tier-$m$ power allocation fraction $\beta_l\in(\theta\sum_{n=0}^{l-1}\beta_n,1)$ holds for all $l\in\{k,\dots,K\}$, then the coverage of the $k$th nearest user to the tier-$m$ BS among the $K$ users, i.e., $\rho_{m,k}$ defined in \eqref{Eqn:DefnCovProbUk}, has a tight lower bound given by
\begin{align}
\rho_{m,k}\gtrapprox \prod_{j=0}^{k-1} \frac{(K-j)}{(K-j)+\sum_{l=1}^{M}\nu_l\ell_{m,l}\left(\vartheta_{k,K}\right)},\,\,\forall k\in\mathcal{K},\label{Eqn:CovProbUserk}
\end{align}
where $\ell_{m,l}(\cdot)$ can be found in \eqref{Eqn:ellfun-mk} and $\vartheta_{k,K}$ is defined as
\begin{align}\label{Eqn:EquivSIRThreshold-mk}
\vartheta_{k,K}\defn \max_{l\in\{k,\ldots,K\}}\left\{\frac{\theta}{\beta_l-\theta\sum_{n=0}^{l-1}\beta_n},0\right\}.
\end{align}
 Moreover, as the user intensity goes to infinity, $\rho_{m,k}$ reduces to the following limit 
\begin{align}
\lim_{\mu\rightarrow\infty}\rho_{m,k}=\prod_{j=0}^{k-1} \frac{(K-j)}{(K-j)+\sum_{l=1}^{M}\ell_{m,l}\left(\vartheta_{k,K}\right)}. \label{Eqn:LowLimitCovProbUserk}
\end{align}
\end{proposition}
\begin{IEEEproof}
See Appendix \ref{App:ProfCovProbNOMA}.
\end{IEEEproof}
The coverage in Proposition \ref{Prop:CovProbNOMA} has some important implications regarding how power allocations among the $K$ NOMA users significantly impact the coverage performance of each user and they are elaborated as follows. First, as shown in the proof of Proposition \ref{Prop:CovProbNOMA}, user $U_k$ cannot decode its own signals almost surely if $\beta_l\leq \theta \sum^{l-1}_{n=0}\beta_n$ for $l\in[k,\ldots, K]$ because the signals of the users farther than the $k$th user cannot be decoded even when there is no interference (\ie decoding the desired signals of each user directly fails due to the NOMA interferences from other $K-1$ users.). Hence, the condition  $\theta \sum^{l-1}_{n=0}\beta_n<\beta_l<1$ for $l\in\mathcal{K}$ is called \textit{``the fundamental constraint on the power allocation for $K$-user (non-coordinated) NOMA transmission"}. As such, $\beta_k$ needs to be properly chosen so that the advantage of NOMA is able to be exploited effectively. Second, $\vartheta_{k,K}$ in \eqref{Eqn:EquivSIRThreshold-mk} indicates $\frac{\theta}{\beta_l-\theta\sum_{n=0}^{l-1}\beta_n}\leq \vartheta_{k,K}$ and $\rho_{m,k}$ must decrease as $k$ increases if $\frac{\theta}{\beta_k-\theta\sum_{n=0}^{k-1}\beta_n}=\vartheta_{k,K}$, which means the farther users might have a worse coverage than the nearer users even though these farther users are allocated more power. For example, if a certain power allocation scheme that allocates enough power to the $K$th (farthest) user gives rise to $\frac{\theta}{\beta_K-\theta\sum_{n=0}^{K-1}\beta_n}=\vartheta_{k,K}$, then all $\vartheta_{k,K}$'s are the same and equal to $\frac{\theta}{\beta_K-\theta\sum_{n=0}^{K-1}\beta_n}$ so that $\rho_{m,k}$ monotonically decreases as $k$ increases. In other words, a user that is farther from its BS has a worse coverage whereas in this case the performance of decoding the signals of the $K$th user dominates the coverage performance of each user.  

\subsection{Throughput Analysis for non-coordinated NOMA}\label{SubSec:RateAnaNonCooPNOMA}
Since the $k$th user associating with BS $X_{m,i}$ needs to successively cancel the interference signals of the $K-k$ users farther than it before decoding its own signals, its link throughput (achievable rate, nats/Hz) based on the result in \eqref{Eqn:DefnCovProbUk} can be defined as
\begin{align}\label{Eqn:DefnLinkThrough-mk}
c_{m,k}\defn\mathbb{E}\left[\log\left(1+\frac{\gamma_{m,k}}{(\sum_{n=0}^{k-1}\beta_n)\gamma_{m,k}/\beta_k+1}\right)\bigg| \gamma_{m,k}\geq\beta_k\vartheta_{k+1,K}\right],
\end{align}
where $k\in\{1,\ldots,K-1\}$ and  $\vartheta_{k+1,K}$ is already defined in \eqref{Eqn:EquivSIRThreshold-mk}. The way of defining $c_{m,k}$ is due to the fact that the achievable link throughput of user $U_k$ should be evaluated whenever user $U_k$ is able to decode the signals of the $K-k$ users farther than it and subtract them from the interference by SIC, \ie the condition $\gamma_{m,k}\geq\beta_k\vartheta_{k+1,K}$ is necessary since it is the condition that user $U_k$ successfully cancels all the signals of the $K-k$ users farther than it (See the explanation for this condition in the proof of Proposition \ref{Prop:CovProbNOMA}). Similarly, for the $K$th user, its link throughput can be defined as
\begin{align}\label{Eqn:DefnLinkThrough-mK}
c_{m,K} \defn\mathbb{E}\left[\log\left(1+\frac{\gamma_{m,K}}{(\sum_{n=0}^{K-1}\beta_n)\gamma_{m,K}/\beta_K+1}\right)\right],
\end{align}
which does not have any condition on $\gamma_{m,K}$ since the $K$th user does not need to cancel any signals of any other NOMA users. The accurate tight lower bounds on $c_{m,k}$ and $c_{m,K}$ are found and shown in the following proposition.
\begin{proposition}\label{Prop:LinkThroughputNonCoopNOMA}
Suppose $\beta_l\in(\theta\sum_{n=0}^{l-1}\beta_n,1)$ holds for all $m\in\mathcal{M}$ and $l\in\{k,\dots,K\}$. For a tier-$m$ BS that is able to arbitrarily schedule $K$ users in its user set, the link throughput of the $k$th nearest user among the $K$ scheduled users is tightly lower bounded by
\begin{align}\label{Eqn:LinkThrouhgput-mk}
c_{m,k}
\gtrapprox &\int_{\vartheta_{k+1,K}}^{\infty}\prod_{j=0}^{k-1}\frac{(K-j)+\sum_{l=1}^{M}\nu_l\ell_{m,l}(\vartheta_{k+1,K})}{(K-j)+\sum_{l=1}^{M}\nu_l\ell_{m,l}(y)}\left[\frac{\beta_k}{(1+y\sum_{n=1}^{k}\beta_n)(1+y\sum_{n=0}^{k-1}\beta_n)}\right]\dif y\nonumber\\
&+\log\left(1+\frac{\beta_k\vartheta_{k+1,K}}{\vartheta_{k+1,K}\sum_{n=0}^{k-1}\beta_n+1}\right),\,k\in\{1,2,\ldots,K-1\}.
\end{align}
Whereas the link throughput of the $K$th user has an accurate tight lower bound given by
\begin{align}\label{Eqn:LinkThroughput-mK}
c_{m,K} 
&\gtrapprox \int_{0}^{\infty}\prod_{j=0}^{K-1}\frac{(K-j)}{(K-j)+\sum_{l=1}^{M}\nu_l\ell_{m,l}(y)} \left(1-\prod_{j=0}^{K-1}\frac{(K-j)+\sum_{l=1}^{M}\nu_l\ell_{m,l}(y)}{(K-j)+\sum_{l=1}^{M}\nu_l\ell_{m,l}(\frac{y}{1-\beta_K})}\right)\frac{\dif y}{(1+y)}.
\end{align}
\end{proposition}
\begin{IEEEproof}
	See Appendix \ref{App:ProofLinkThroughputNonCoopNOMA}.
\end{IEEEproof}

\begin{remark}\label{Rem:AsyLinkThroughput1}
When $K\rightarrow\infty$, we have $\beta_k\rightarrow 0$ and $\vartheta_{k+1,K}\rightarrow\infty$ for all $k\in\mathcal{K}$. In this case, 
the result in  \eqref{Eqn:LinkThrouhgput-mk} asymptotically reduces to $c_{m,k}\gtrapprox \log(1+\beta_k/\sum_{n=0}^{k-1}\beta_n+\vartheta^{-1}_{k+1,K})$ whereas the result in \eqref{Eqn:LinkThroughput-mK} asymptotically approaches to zero. These asymptotic results indicate that the SIR of the farthest user is very small and other nearer $K-1$ users cannot improve their link throughputs too much by performing SIC. The sum throughput of the NOMA users would be largely degraded if too many users are scheduled. Hence, choosing an appropriate number of the scheduled users for NOMA transmission is important. 
\end{remark}

The results in \eqref{Eqn:LinkThrouhgput-mk} and \eqref{Eqn:LinkThroughput-mK} present a very disparate nature in throughput owing to SIC, \ie the $k$th user  that successfully performs SIC can  achieve a link throughput no less than the first term at the right side of \eqref{Eqn:LinkThrouhgput-mk} whereas the $K$th user that does not perform SIC cannot achieve a non-zero minimum link throughput. This indicates that $c_{m,k}$'s and $c_{m,K}$ all augment as long as a power allocation scheme is able to increase $\beta_k\vartheta_{k+1,K}$ and $\beta_K$ simultaneously. Also, the link throughput results in \eqref{Eqn:LinkThrouhgput-mk} and \eqref{Eqn:LinkThroughput-mK} both imply the following asymptotic result:
\begin{align}\label{Eqn:LinkRateSoleUser}
\lim_{\beta_k\rightarrow 1}c_{m,k}=\mathbb{E}\left[\log\left(1+\frac{\gamma_{m,k}}{\beta_k}\right)\right]\gtrapprox \int_{0}^{\infty} \frac{\dif y}{(1+y)\left[1+\sum_{l=1}^{M}\nu_l\ell_{m,l}(y)\right]},
\end{align}
which is exactly the tight lower bound on the link throughput of a tier-$m$ BS serving a sole user and it is the upper bound on $c_{m,k}$ for all $k\in\mathcal{K}$ since the full transmit power is only allocated to a user. This obviously means that the results in \eqref{Eqn:LinkThrouhgput-mk} and \eqref{Eqn:LinkThroughput-mK} all reduce to the link throughput of a single user as $\beta_k$ goes to one. More importantly, using the results in \eqref{Eqn:ProofLinkThrputUserk} and \eqref{Eqn:ProofLinkThrputUserK2} in Appendix  \ref{App:ProofLinkThroughputNonCoopNOMA} we are able to show that the sum throughput of the $K$ NOMA users is strictly larger than the link throughput of a single user in \eqref{Eqn:LinkRateSoleUser}, that is,
\begin{align}
\sum_{k=1}^{K} c_{m,k}>\int_{0}^{\infty} \frac{\dif y}{(1+y)\left[1+\sum_{l=1}^{M}\nu_l\ell_{m,l}(y)\right]},
\end{align}  
and this manifests the fact that\textit{ the NOMA scheme is always able to achieve higher throughput than the OMA schemes as long as SIC performs well}. Although this fact somewhat may not be very surprised, to the best of our knowledge it is firstly shown here for a large-scale HetNet model with fading channels. In addition, note that all $c_{m,k}$'s asymptotically reduce to their lowest limits that are equal to the lower bound in \eqref{Eqn:LinkThrouhgput-mk} with $\nu_l=1$ for all $l\in\mathcal{M}$ as the user intensity goes to infinity since no void BSs exist in the network in this scenario.

\section{Downlink Coverage  and Throughput Analysis for coordinated JT-NOMA}\label{Sec:CoopNOMA}
In Section \ref{Sec:NonCoopNOMA}, the coverage probability and link throughput of the users for the non-coordinated NOMA scheme are investigated and shown to be severely impacted by the power allocation scheme among NOMA users as well as the SIC performance. To alleviate the impact on the coverage and link throughput due to imperfect power allocation as well as SIC, in this section we propose a coordinated JT-NOMA scheme that is able to significantly improve the coverage and link throughput of all the NOMA users. The fundamental idea behind this coordinated JT-NOMA scheme is to coordinate some void BSs to jointly transmit the signals of the furthest user among the $K$ users scheduled by a tier-$m$ BS in that enhancing the signal strength of the $K$th user benefits the performance of decoding the signals of all these users. For this proposed coordinated JT-NOMA scheme, in the following we will analyze the coverage and the link throughput for each of the $K$ users scheduled by a tier-$m$ BS. According to the analytical results, we will be able to see how much the NOMA transmission performance can be improved by the proposed coordinated JT-NOMA scheme in terms of the average cell coverage and cell throughput of each BS.

\subsection{Coverage Analysis for coordinated JT-NOMA}\label{SubSec:CovAnaCooPNOMA}
 
By following the similar analytical approach in Section \ref{SubSec:CovAnaNonCooPNOMA}, consider BS $X_{m,i}$ located at the origin and it is able to arbitrarily schedule $K$ users among all its tagged users for downlink NOMA transmission. Also, we assume that all void BSs are coordinated to jointly transmit the signals of the farthest user (the $K$th user)\footnote{The reason of making such an assumption that ``all'' void BSs can be coordinated to do joint transmission is two-fold: First, such an assumption leads to very much tractable analyses in the coverage and link throughput, as shown in our previous work in \cite{CHLPCC17}. Second, under this assumption we can study the fundamental limits on the coverage and link throughput achieved by the proposed coordinated JT-NOMA scheme. Note that all coordinated void BSs do not need to know the channel ordering status between the $K$ scheduled users. They just need to know the signals of the farthest user, which can be accomplished by the BS coordination techniques.}. Since all the void BSs are coordinated to transmit the signals of the $K$th user, the coverage probability of the $k$th user for the coordinated JT-NOMA scheme, based on the desired SIR $\gamma_{m,k}$ defined in \eqref{Eqn:DSIR}, can be defined as
 \begin{align}
\rho_{m,k}&\defn\mathbb{P}\left[\frac{\beta_kP_mH_{m,i,k}\|U_k\|^{-\alpha}}{(\sum_{n=0}^{k-1}\beta_n)P_m\frac{H_{m,i,k}}{\|U_k\|^{\alpha}}+I_{m,k}}\geq\theta,\ldots,\frac{\beta_KP_mH_{m,i,k}\|U_k\|^{-\alpha}+S_{m,k}}{(\sum_{n=0}^{K-1}\beta_n)P_m\frac{H_{m,i,k}}{\|U_k\|^{\alpha}}+I_{m,k}}\geq\theta\right]\nonumber\\
&=\mathbb{P}\left[\frac{\gamma_{m,k}}{(\sum_{n=0}^{k-1}\beta_n)\gamma_{m,k}/\beta_k+1}\geq\theta, \ldots, \frac{\beta_K\gamma_{m,k}/\beta_k+S_{m,k}/I_{m,k}}{(\sum_{n=0}^{K-1}\beta_n)\gamma_{m,k}/\beta_k+1}\geq \theta\right],\label{Eqn:DefnCovProbCoopNOMA}
 \end{align} 
where $S_{m,k}\defn  \sum_{l,j:X_{l,j}\in\Phi} (1-V_{l,j})P_lH_{l,j,k} \|X_{l,j}-U_k\|^{-\alpha}$ denotes the sum of the signal powers of user $U_K\in\mathcal{C}_{m,i}$ coming from all coordinated void BSs that are used to jointly transmit the signals of $U_K$. The tight lower bound on $\rho_{m,k}$ in \eqref{Eqn:DefnCovProbCoopNOMA} is derived and shown in the following.

\begin{proposition}\label{Prop:CovProbCooPNOMA}
If the coordinated JT-NOMA scheme is performed in the HetNet, the coverage $\rho_{m,k}$ of the $k$th user in \eqref{Eqn:DefnCovProbCoopNOMA} with the power allocation constraint on  $\beta_l\in(\theta\sum_{n=0}^{l-1}\beta_n,\beta_K-\theta\sum_{n=l}^{K-1}\beta_n)$ with $\beta_K\in(\theta\sum_{n=0}^{K-1}\beta_n,1)$ for all $l\in\{k,\ldots,K-1\}$ is tightly lower bounded by 
\begin{align}\label{Eqn:CoverProbCoopNOMAUserk}
\rho_{m,k}\gtrapprox \prod_{j=0}^{k-1}\frac{(K-j)}{(K-j)+\sum_{l=1}^{K}\nu_l\ell_{m,l}\left(\vartheta_{k,K-1}\right)},\,\,k\in\{1,\ldots,K-1\}
\end{align}
where $\ell_{m,l}(\cdot)$ and $\vartheta_{k,K-1}$ are already defined in \eqref{Eqn:ellfun-mk} and \eqref{Eqn:EquivSIRThreshold-mk}, respectively. The coverage of the $K$th user can be accurately approximated by
\begin{align}\label{Eqn:CoverProbCoopNOMAUserK}
\rho_{m,K}\approx \prod_{j=0}^{K-1} \frac{(K-j)}{(K-j)+\left[\sum_{l=1}^{M}\nu_l\ell_{m,l}\left(\vartheta_{K,K}\right)+(1-\nu_l)\widetilde{\ell}_{m,l}\left(\frac{\vartheta_{K,K}}{\theta}\right)\right]^+},
\end{align}  
where $(x)^+\defn\max\{y,0\}$ and $\widetilde{\ell}_{m,l}(\cdot)$ is defined as
\begin{align}
\widetilde{\ell}_{m,l}(x)\defn \varphi_l\left(\frac{x \omega_mP_l}{\omega_lP_m}\right)^{\frac{2}{\alpha}}\int_{\left(\frac{\omega_lP_m}{x\omega_m P_l}\right)^{\frac{2}{\alpha}}}^{\infty}\mathbb{E}\left[1-e^{t^{-\frac{\alpha}{2}}H}\right]\dif t,
\end{align}
where $H\sim\text{Exp}(1)$. Furthermore, we have the following asymptotic result of $\rho_{m,k}$ 
\begin{align}\label{Eqn:CoverProbCoopNOMAUserkNoVoid}
\lim_{\mu\rightarrow\infty}\rho_{m,k}= \prod_{j=0}^{k-1}\frac{(K-j)}{(K-j)+\sum_{l=1}^{M}\ell_{m,l}\left(\vartheta_{k,K}\right)},\,\,k\in\{1,\ldots,K\}
\end{align}
as the user intensity $\mu$ goes to infinity.
\end{proposition}
\begin{IEEEproof}
See Appendix \ref{App:ProfCovProbCooPNOMA}.
\end{IEEEproof}

The coverage results in Proposition \ref{Prop:CovProbCooPNOMA} clearly indicate how the coordinated JT-NOMA scheme improves the coverage probabilities of the $K$ users. If we compare \eqref{Eqn:CoverProbCoopNOMAUserk} with \eqref{Eqn:CovProbUserk} for $k\in\{1,\ldots,K-1\}$, we can see the coverage probability of the $k$th user in \eqref{Eqn:CoverProbCoopNOMAUserk} is higher than that in \eqref{Eqn:CovProbUserk} since $\ell_{m,l}(x)$ is a monotonically increasing function of $x$ and $\vartheta_{k,K-1}$ in \eqref{Eqn:CoverProbCoopNOMAUserk} cannot be greater than  $\vartheta_{k,K}$ in \eqref{Eqn:CovProbUserk}. This is because coordinated JT-NOMA with appropriate power allocation schemes is able to make the SIR of the $K$th user higher than the SIRs of the other $K-1$ users. Thus, the coverage probability of the first $K-1$ users does not depend on the SIR performance of the $K$th user. Obviously, the coverage probability of the $K$th user is enhanced as well if comparing \eqref{Eqn:CoverProbCoopNOMAUserK} with \eqref{Eqn:CovProbUserk} for $k=K$ since the term $\widetilde{\ell}_{m,l}(\vartheta_{K,K}/\theta)$ in \eqref{Eqn:CoverProbCoopNOMAUserK} is negative and it is not in \eqref{Eqn:CovProbUserk} for $k=K$. Note that the coverage probabilities of the $K$ users achieved by coordinated JT-NOMA increase as the user intensity reduces since more void BSs can be coordinated to improve the signal strength of the $K$th user. On the contrary, the coverage performance of coordinated JT-NOMA degrades as the user intensity increases. Thus, an interesting and important problem that can be further studied is about how to maintain an appropriate cell load of the BSs in each tier (see \eqref{Eqn:pmfNumUsers}) so that there exists a good number of the void BSs that can be coordinated to perform the proposed coordinated JT-NOMA scheme for a given user intensity. 

Note that the power allocation in Proposition \ref{Prop:CovProbCooPNOMA} has a sticker constraint on $\beta_l$, (\ie $\beta_l\in(\theta\sum_{n=0}^{l-1}\beta_n,\beta_K-\theta\sum_{n=l}^{K-1}\beta_l)$) than that on $\beta_{l}$ in Proposition \ref{Prop:CovProbNOMA}. This sticker constraint on $\beta_m$ is obtained by facilitating the derivations of the coverage probabilities in the proof of Proposition \ref{Prop:CovProbCooPNOMA} when the coordinated JT-NOMA scheme is adopted. In fact, the coverage probabilities essentially can be improved by coordinated JT-NOMA for any $\beta_l\in(\theta\sum_{n=0}^{l-1}\beta_n,1)$. Most importantly, this constraint $\beta_l\in(\theta\sum_{n=0}^{l-1}\beta_n,\beta_K-\theta\sum_{n=l}^{K-1}\beta_l)$ lets us realize that coordinated JT-NOMA can achieve a higher coverage of each user with a less power allocated to the first $K-1$ users. 

\subsection{Throughput Analysis for coordinated JT-NOMA} \label{SubSec:RateAnaCooPNOMA}
In this subsection, we turn our attention on the downlink throughput achieved by a tier-$m$ BS for the proposed coordinated JT-NOMA scheme. Since all void BSs are assumed to jointly transmit the signals of the $K$th NOMA user of the tier-$m$ BS only, the link throughput of the $k$th user for $ k\in\{1,\dots,K-2\}$ in this case based on \eqref{Eqn:DefnCovProbCoopNOMA} can be defined as
\begin{align}\label{Eqn:DefnLinkThroughmkCoopNOMA}
c_{m,k} \defn\mathbb{E}\left[\log\left(1+\frac{\gamma_{m,k}}{(\sum_{n=0}^{k-1}\beta_n)\frac{\gamma_{m,k}}{\beta_k}+1}\right)\bigg| \gamma_{m,k}\geq\beta_k\vartheta_{k+1,K-1}\right].
\end{align}
For the $K-1$th user, its link throughput is defined as
\begin{align}
c_{m,K-1}\defn \mathbb{E}\left[\log\left(1+\frac{\gamma_{m,K-1}}{(\sum_{n=0}^{K-2}\beta_n)\frac{\gamma_{m,K-1}}{\beta_{K-1}}+1}\right)\bigg| \gamma_{m,K-1}\geq\frac{\beta_{K-1}\theta-\frac{S_{m,K-1}}{I_{m,K-1}}}{\beta_K-\theta\sum_{n=0}^{K-1}\beta_n}\right].
\end{align} 
These two definitions are based on the same idea of defining $c_{m,k}$ in \eqref{Eqn:DefnLinkThrough-mk} for non-coordinated NOMA and the proof of Proposition \ref{Prop:CovProbCooPNOMA}. Whereas the link throughput of the $K$th user can be defined as
\begin{align}\label{Eqn:DefnLinkThroughmKCoopNOMA}
c_{m,K}\defn\mathbb{E}\left[\log\left(1+\frac{\gamma_{m,K}+\frac{S_{m,K}}{I_{m,K}}}{(\sum_{n=0}^{K-1}\beta_n)\frac{\gamma_{m,K}}{\beta_K}+1}\right)\right]
\end{align}
based on the resulting SIR of the $K$th user in \eqref{Eqn:DefnCovProbCoopNOMA}. The approximated and accurate results of $c_{m,k}$ and $c_{m,K}$ are found in the following proposition. 
\begin{proposition}\label{Prop:LinkThoughput_CoopNOMA}
Let  $\beta_l\in(\theta\sum_{n=0}^{l-1}\beta_n,\beta_K-\theta\sum_{n=l}^{K-1}\beta_n)$ with $\beta_K\in(\theta\sum_{n=0}^{K-1}\beta_n,1)$ hold for all $m\in\mathcal{M}$, $l\in[k,\ldots,K-1]$ and $k\in\{1,\ldots, K-1\}$. The tier-$m$ link throughput of the $k$th user in \eqref{Eqn:DefnLinkThroughmkCoopNOMA} for $k\in\{1,\ldots,K-2\}$ has a tight lower bound given by
\begin{align}\label{Eqn:LinkThrough-mkCoopNOMA1}
c_{m,k}\gtrapprox &\int_{\vartheta_{k+1,K-1}}^{\infty}\prod_{j=0}^{k-1}\frac{(K-j-1)+\sum_{l=1}^{M}\nu_l\ell_{m,l}(\vartheta_{k+1,K-1})}{(K-j-1)+\sum_{l=1}^{M}\nu_l\ell_{m,l}(y)}\left[\frac{\beta_k}{(1+y\sum_{n=1}^{k}\beta_n)(1+y\sum_{n=0}^{k-1}\beta_n)}\right]\dif y\nonumber\\
&+\log\left(1+\frac{\beta_k\vartheta_{k+1,K-1}}{\vartheta_{k+1,K-1}\sum_{n=0}^{k-1}\beta_n+1}\right).
\end{align}
Whereas the link throughput of the $(K-1)$th user has a tight lower bound given by
\begin{align}\label{Eqn:LinkThrough-mkCoopNOMA2}
c_{m,K-1}\gtrapprox & \int_{0}^{\infty}\left(1-\prod_{j=0}^{K-2}\frac{(K-j-1)+\sum_{l=1}^{M}\nu_l\ell_{m,l}(y)}{(K-j-1)+\sum_{l=1}^{M}\nu_l\ell_{m,l}(\frac{y}{1-\beta_{K-1}})}\right)\nonumber\\
&\times \left(\prod_{j=0}^{K-2}\frac{(K-j-1)}{(K-j-1)+\sum_{l=1}^{M}\nu_l\ell_{m,l}(y)}\right)\frac{\dif y}{(1+y)}.
\end{align}
For the $K$th user, its link throughput is given by
\begin{align}\label{Eqn:LinkThrough-mkCoopNOMA3}
c_{m,K}\approx & \int_{0}^{\frac{\beta_K}{\sum_{l=0}^{K-1}\beta_l}}\prod_{j=0}^{K-1} \frac{(K-j)}{(K-j)+\left[\sum_{l=1}^{M}\nu_l\ell_{m,l}\left(y_K\right)+(1-\nu_l)\widetilde{\ell}_{m,l}\left(\frac{y_{K}}{\theta}\right)\right]^+}\frac{\dif y}{ (1+y)}.
\end{align}
\end{proposition}
\begin{IEEEproof}
See Appendix \ref{App:ProofLinkThroughputCoopNOMA}.
\end{IEEEproof}

\begin{remark}\label{Rem:AsyLinkThroughput2}
When $K\rightarrow\infty$, we have $\beta_k\rightarrow 0$ and $\vartheta_{k+1,K-1}\rightarrow\infty$ for all $k\in\mathcal{K}$. In this case, 
the result in  \eqref{Eqn:LinkThrough-mkCoopNOMA1}  asymptotically reduces to $c_{m,k}\gtrapprox \log(1+\beta_k/\sum_{n=0}^{k-1}\beta_n+\vartheta^{-1}_{k+1,K-1})$ for $k\in\{1,\ldots, K-2\}$ whereas the results in \eqref{Eqn:LinkThrough-mkCoopNOMA2} and \eqref{Eqn:LinkThrough-mkCoopNOMA3} asymptotically approach to zero. This is similar to the fact pointed out in Remark \ref{Rem:AsyLinkThroughput1} that scheduling too many users at the same time would significantly degrade the sum throughput of the users even in the case of the coordinated JT-NOMA scheme.
\end{remark}

According to the link throughput results found in Proposition \ref{Prop:LinkThoughput_CoopNOMA}, we can easily realize that the coordinated JT-NOMA scheme indeed improves the link throughput of each NOMA user since the coordinated void BSs directly enhance the SIR of the $K$th user and make the SIR performance of the other $K-1$ users not significantly impacted by the SIR of the $K$th (because the SIC performance of the $K-1$ users is significantly improved by the coordinated void BSs). As a result, when there are a large number of viod BSs in the network and coordinated JT-NOMA is used, we can reduce the power allocation of the $K$th user so that the rest of the $K-1$ users can acquire more power so as to improve their coverage and throughput. Furthermore, the results found in Proposition \ref{Prop:LinkThoughput_CoopNOMA} are the throughput limits achieved by arbitrarily scheduling $K$ NOMA users and coordinating all void BSs so that they highly depend on the user and BS intensities and they all reduce to their corresponding results in Proposition \ref{Prop:LinkThroughputNonCoopNOMA} as the user intensity goes to infinity.  

\section{Optimal Power Allocation Analysis}\label{Sec:OptimalPowerAllocation}
In Sections \ref{Sec:NonCoopNOMA} and \ref{Sec:CoopNOMA}, we have analyzed the coverage probability and link throughput of a user for the non-coordinated and coordinated JT-NOMA schemes and pointed out that an appropriate power allocation scheme for the $K$ NOMA users significantly benefits the coverage and throughput performances of these users. In the following, we will investigate how to optimally allocate transmit powers to the $K$ users in order to maximize the (average) cell coverage and cell throughput.

\subsection{Optimal Power Allocation for Maximizing Cell Coverage}
Let vector $\mathbf{v}_{\beta}\defn[\beta_{1},\ldots,\beta_K]^{\mathrm{T}}\in[0,1]^K$ (where $\mathrm{T}$ denotes the ``transpose'' operator) be a $K\times 1$ power allocation vector for the $K$ users associating with a tier-$m$ BS and we can formulate the optimization problem of $\mathbf{v}_{\beta}$ in order to maximize the average of the coverage probabilities in a tier-$m$ cell with a given $\theta>0$ as follows
\begin{align}\label{Eqn:OptCovProb}
\begin{cases}
\text{max}_{\mathbf{v}_{\beta}} &\frac{1}{K}\sum_{k=1}^{K}\rho_{m,k}\\
\text{s.t. } &\mathbf{v}_{\beta}\in\mathcal{V}_{\beta}(\theta)
\end{cases},
\end{align}
in which the objective function is called \textbf{the (average) tier-$m$ cell coverage} for $K$-user NOMA and $\mathcal{V}_{\beta}(\theta)$ is the feasible set of power allocation vector $\mathbf{v}_{\beta}$ for a given $\theta$. 

Since the explicit results of $\rho_{m,k}$ have been found in Propositions \ref{Prop:CovProbNOMA} and \ref{Prop:CovProbCooPNOMA}, there exists an optimal power allocation vector $\mathbf{v}^{\star}_{\beta}\defn[\beta^{\star}_1,\ldots,\beta^{\star}_K]^{\mathrm{T}}$ that maximizes the tier-$m$ cell coverage in \eqref{Eqn:OptCovProb}, as shown in the following proposition.
\begin{proposition}\label{Prop:OptimalPowerAllocationCoverage}
For the non-coordinated NOMA scheme, the optimization problem in \eqref{Eqn:OptCovProb} with the feasible set $\mathcal{V}_{\beta}(\theta)$ for a given $\theta>0$ given by
\begin{align}\label{Eqn:PowerAllSetNonCoopNOMA}
\mathcal{V}_{\beta}(\theta)\defn\left\{\mathbf{v}_{\beta}\in[0,1]^K:\sum_{k=1}^{K}\beta_k=1,   0<\theta\sum^{l-1}_{k=1}\beta_k<\beta_l\leq 1,l\in\mathcal{K}\right\},
\end{align}
has an optimal vector $\mathbf{v}^{\star}_{\beta}\in\mathcal{V}_{\beta}(\theta)$ that maximizes the tier-$m$ cell coverage in \eqref{Eqn:OptCovProb}. Similarly, for the coordinated JT-NOMA scheme,  the following set
\begin{align}\label{Eqn:PowerAllSetCoopNOMA}
\mathcal{V}_{\beta}(\theta)\defn\left\{\mathbf{v}_{\beta}\in[0,1]^K:\sum_{k=1}^{K}\beta_k=1,    0< \beta_l+\theta\sum^{K-1}_{k=l}\beta_k< \beta_K\leq 1, l\in\mathcal{K}\right\},
\end{align}
is a feasible set of the optimization problem in \eqref{Eqn:OptCovProb} and there exists an optimal vector $\mathbf{v}_{\beta}\in\mathcal{V}_{\beta}$ that maximizes the tier-$m$ cell coverage. 
\end{proposition}
\begin{IEEEproof}
See Appendix \ref{App:ProofOptimalPowerAllocationCoverage}.
\end{IEEEproof}  
Accordingly, an optimal power allocation vector for the $K$ NOMA users indeed exists in the feasible set specified in \eqref{Eqn:PowerAllSetCoopNOMA}. Due to the complexity of $\rho_{m,k}$ for $\mathbf{v}_{\beta}$, the optimization problem in \eqref{Eqn:OptCovProb} in general is not convex and thus $\mathbf{v}^{\star}_{\beta}$ may not be unique. Nevertheless, the optimal power allocation vector can be numerically found by some existing heuristic algorithms (such as genetic algorithms, simulated annealing algorithms and ant colony algorithms, etc. \cite{IBJLPS13}) once $\theta$ and other necessary parameters in \eqref{Eqn:PowerAllSetCoopNOMA} are designated.  Note that for some special case with a small number of users, such as $K=2$, a unique optimal $\mathbf{v}^{\star}_{\beta}$ can be found (This will be numerically verified in Section \ref{Sec:Simulation}.). Note that the upper bound on $\theta$ for non-coordinated NOMA is $\theta <\min_{l\in\{k,\ldots,K\}}\{\beta_l/\sum_{n=0}^{l-1}\beta_n\}$ by inferring from the $K-k+1$ conditions $\theta \sum_{n=0}^{l-1}\beta_n<\beta_l$ for all $l\in\{k,\ldots,K\}$, whereas the upper bound on $\theta$ for coordinated JT-NOMA is $(\beta_K-\beta_l)/\sum_{k=l}^{K}\beta_k$.  Once $\theta$ is determined, these two bounds pose a constraint on power allocation of performing NOMA. Moreover, we can expect that $\mathbf{v}_{\beta}^{\star}$ for coordinated JT-NOMA may be element-wisely higher than that for non-coordinated NOMA since joint transmission enhances the signal power of the $K$th user so that allocating less power to the $K$th user and more power to the other $K-1$ users would not degrade the optimal value of the tier-$m$ cell coverage. 

\subsection{Optimal Power Allocation for Maximizing Cell Throughput}
Since the explicit expressions of the link throughputs of the $K$ NOMA users are already found in Propositions \ref{Prop:LinkThroughputNonCoopNOMA} and \ref{Prop:LinkThoughput_CoopNOMA}, we also can formulate an optimization problem of power allocation that maximizes the sum link throughput of a tier-$m$ BS serving $K$ NOMA users as follows
\begin{align}\label{Eqn:SumRateOptimizationNonCoopNOMA}
\begin{cases}
\max_{\mathbf{v}_{\beta}} &\sum_{k=1}^{K} c_{m,k} \\
\text{s.t. } & \mathbf{v}_{\beta}\in\mathcal{V}_{\beta}(\theta)
\end{cases},
\end{align}
where the objective function is called \textbf{the tier-$m$ cell throughput} of $K$ NOMA users. The optimal power allocation vector $\mathbf{v}_{\beta}$ for \eqref{Eqn:SumRateOptimizationNonCoopNOMA} as stated in the following proposition. 
\begin{proposition}\label{Prop:OptPowerAllSumRate}
For the non-coordinated NOMA scheme, the optimization problem in \eqref{Eqn:SumRateOptimizationNonCoopNOMA} with the feasible set $\mathcal{V}_{\beta}(\theta)$ defined in \eqref{Eqn:PowerAllSetNonCoopNOMA} has an optimal solution $\mathbf{v}^{\star}_{\beta}\in\mathcal{V}_{\beta}(\theta)$ that maximizes the tier-$m$ cell throughput\footnote{The optimal power allocation vector found in this proposition may be different from that found in Proposition \ref{Prop:OptimalPowerAllocationCoverage} since the two optimization problems in \eqref{Eqn:OptCovProb} and \eqref{Eqn:SumRateOptimizationNonCoopNOMA} have distinct objective functions.}. Likewise,  for the coordinated JT-NOMA scheme, an optimal vector $\mathbf{v}_{\beta}$ that is able to maximize the tier-$m$ cell throughput can be found in set $\mathcal{V}_{\beta}(\theta)$ defined in \eqref{Eqn:PowerAllSetCoopNOMA}.
\end{proposition}
\begin{IEEEproof}
The proof is omitted here since it is similar to the proof of Proposition \ref{Prop:OptimalPowerAllocationCoverage}.
\end{IEEEproof}
\noindent Generally speaking, the optimization problem in \eqref{Eqn:SumRateOptimizationNonCoopNOMA} is not convex as well and its optimal solution can only be found by numerical techniques due to the complicate expression of $c_{m,k}$. However, similar to the case of the tier-$m$ coverage with a small number of the NOMA users, the optimal solution to \eqref{Eqn:SumRateOptimizationNonCoopNOMA} is analytically much tractable and might be found uniquely. Finally, it is worth pointing out that the optimal value of the tier-$m$ cell throughput must be greater than the link throughput of a sole user since the sum of the link throughputs for any power allocations is no less than the link throughput of a sole user as indicated in \eqref{Eqn:LinkRateSoleUser}. We will validate this issue by numerical simulations in Section \ref{Sec:Simulation}.  

\section{Numerical Results}\label{Sec:Simulation}
Some numerical results are provided in this section to validate the coverage and link throughput analyses with non-coordinated and coordinated JT-NOMA schemes in the previous sections. Here we consider a two-tier HetNet consisting a tier of macrocell BSs and a tier of picocell BSs. Each BS can at most schedule two NOMA users, \ie $K=2$. The network parameters for simulation are listed in Table \ref{Tab:SimPara}. We first present the numerical results of the coverage and link throughput with a specific power allocation between the two NOMA users and then we present how the numerical results of the cell coverages and the cell throughputs change with the power allocations between the two users.
\begin{table}[!t]
\centering
\caption{Network Parameters for Simulation}\label{Tab:SimPara}
\begin{tabular}{|c|c|c|}
\hline Parameter $\setminus$ BS Type (Tier $m$)& Macrocell BS (1) & Picocell BS (2)\\ 
\hline Power $P_m$ (W) & 20 & 5\\
\hline User Intensity $\mu$ (users/m$^2$) &\multicolumn{2}{c|}{$5\times 10^{-4}$} \\ 
\hline Intensity $\lambda_m$ (BSs/m$^2$) & $1.0\times 10^{-6}$ & $[\frac{\mu}{3}, 2\mu]$    \\ 
\hline Number of NOMA Users Scheduled $K$ &\multicolumn{2}{c|}{2} \\ 
\hline Power Allocation Vector $\mathbf{v}_{\beta}$ (if applicable) & \multicolumn{2}{c|}{$[\frac{1}{4}\,\,, \frac{3}{4}]^{\mathrm{T}}$} \\
\hline SIR Threshold $\theta$ &\multicolumn{2}{c|}{1} \\ 
\hline Pathloss Exponent $\alpha$ & \multicolumn{2}{c|}{4} \\  
\hline User Association Bias $\omega_m$ (Nearest BS Association) & \multicolumn{2}{c|}{1}  \\ 
\hline 
\end{tabular} 
\end{table}

\subsection{Numerical Results for Coverage and Link Throughput}
\begin{figure}[t!]
	\centering
	\includegraphics[scale=0.3]{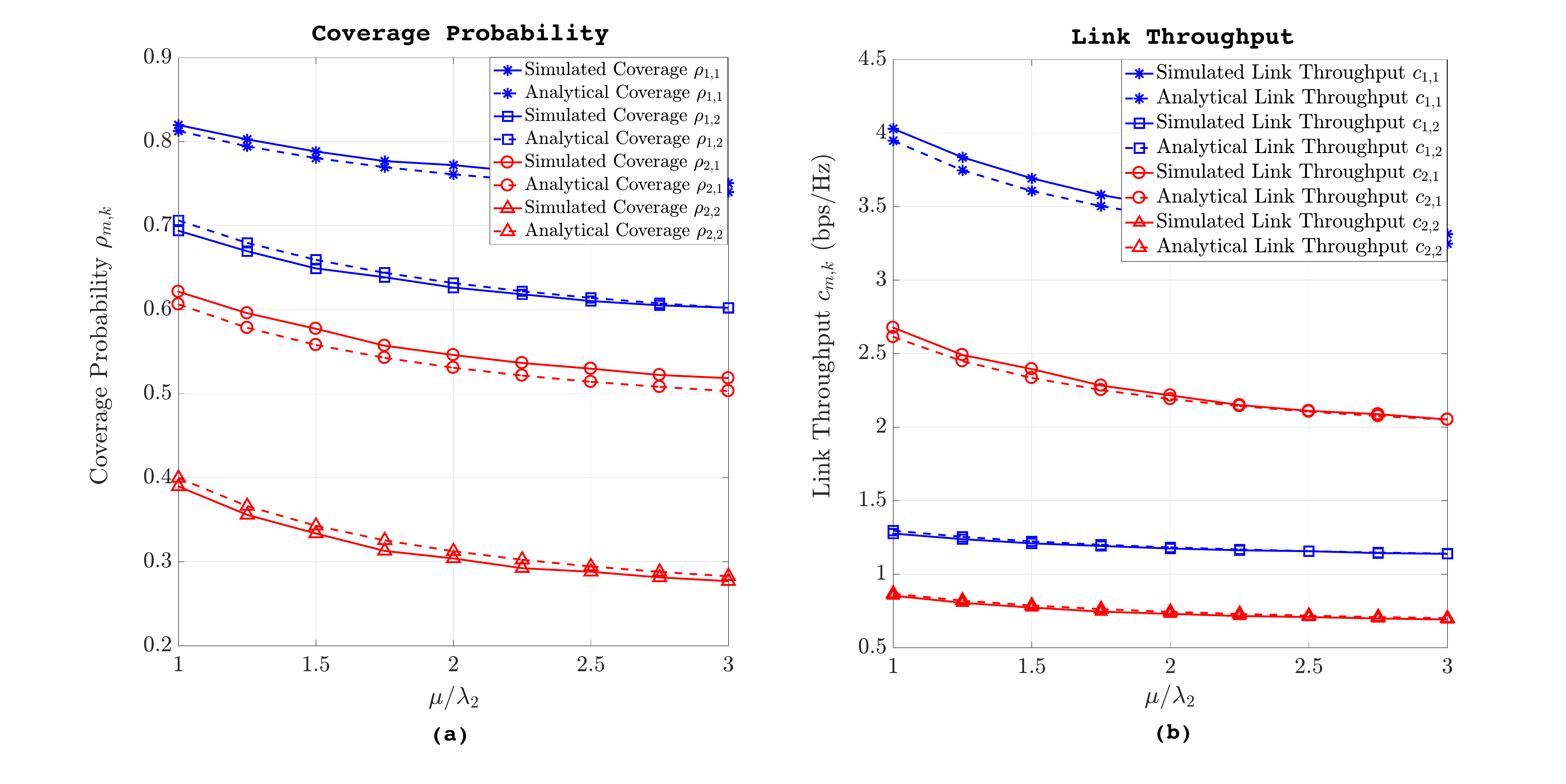}
	\caption{Numerical results of the non-coordinated NOMA scheme with a fixed power allocation $\mathbf{v}_{\beta}=[\frac{1}{4}\,\frac{3}{4}]^{\mathrm{T}}$: (a) Coverage Probability, (b) Link Throughput.}
	\label{Fig:CovThrNonCoopNOMAFixedPowerAll}
\end{figure}

\begin{figure}[t!]
	\centering
	\includegraphics[scale=0.3]{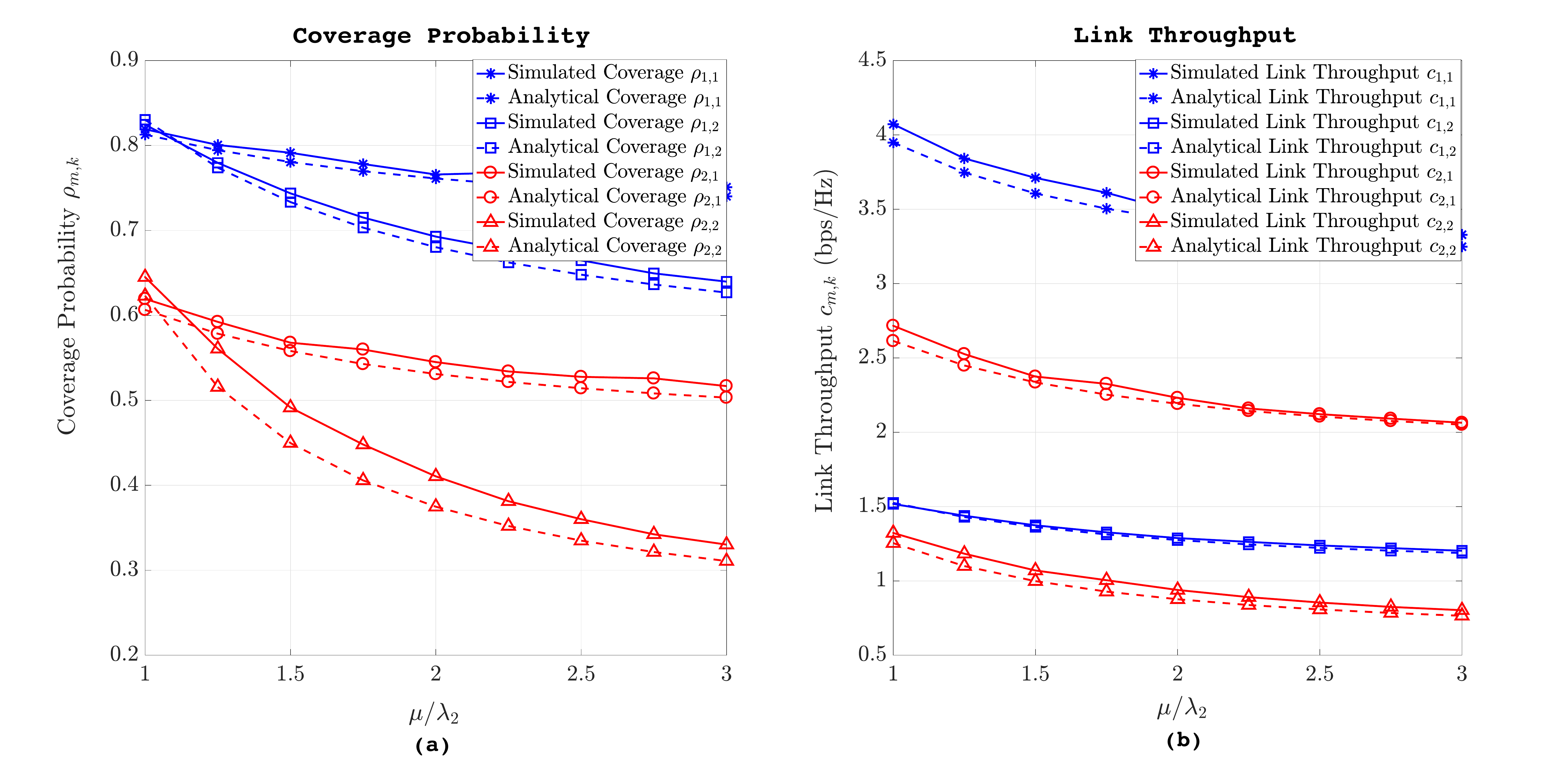}
	\caption{Numerical results of the coordinated JT-NOMA scheme with a fixed power allocation $\mathbf{v}_{\beta}=[\frac{1}{4}\,\frac{3}{4}]^{\mathrm{T}}$: (a) Coverage Probability, (b) Link Throughput.}
	\label{Fig:CovThrCoopNOMAFixedPowerAll}
\end{figure}

Fig. \ref{Fig:CovThrNonCoopNOMAFixedPowerAll} shows the simulation results of the coverage probabilities and link throughputs of the two users for the non-coordinated NOMA scheme.
As can be seen in the figure, all the analytical results are pretty close to their corresponding simulated results, which validates the correctness and accuracy of our previous analyses. Also, we can see that the coverage probabilities of the user in picocells are significantly smaller than those in the marcocells owing to the large transmit power of the marcocell BSs. The coverage probabilities and link throughputs essentially decrease as the user intensity increases since the interference increases due to the increase in the non-void probability and thus the intensity of the non-void BSs in the network increases. Accordingly, all coverage probabilities and link throughputs eventually coverage to a constant value as the user intensity goes to infinity. The simulation results of the coordinated JT-NOMA scheme are shown in Fig. \ref{Fig:CovThrCoopNOMAFixedPowerAll} and we also can see that all analytical results are very close to their corresponding simulated results. In addition, all results in Fig. \ref{Fig:CovThrCoopNOMAFixedPowerAll} are better than those in Fig. \ref{Fig:CovThrNonCoopNOMAFixedPowerAll}, especially for the users associating with a picocell BS. Thus, coordinated JT-NOMA indeed improves the SIR performance of all users. For example, for $\mu/\lambda_2\approx 1.25$, $\rho_{2,2}$ in Fig. \ref{Fig:CovThrCoopNOMAFixedPowerAll} is about $56\%$ higher than $\rho_{2,2}$ in Fig. \ref{Fig:CovThrNonCoopNOMAFixedPowerAll}. Note that the coverage and link throughput performances of the users associating with a macrocell BS seem not improved very much by coordinated JT-NOMA and this is because the transmit power of the macrocell BSs is much higher than that of the picocell BSs and the intensity of the macrocell BSs is much smaller than that of the picocell BSs (\ie $\lambda_2\gg\lambda_l$). 

\subsection{Numerical Results for Cell Coverage and Cell Throughput}
\begin{figure}[t!]
	\centering
	\includegraphics[scale=0.35]{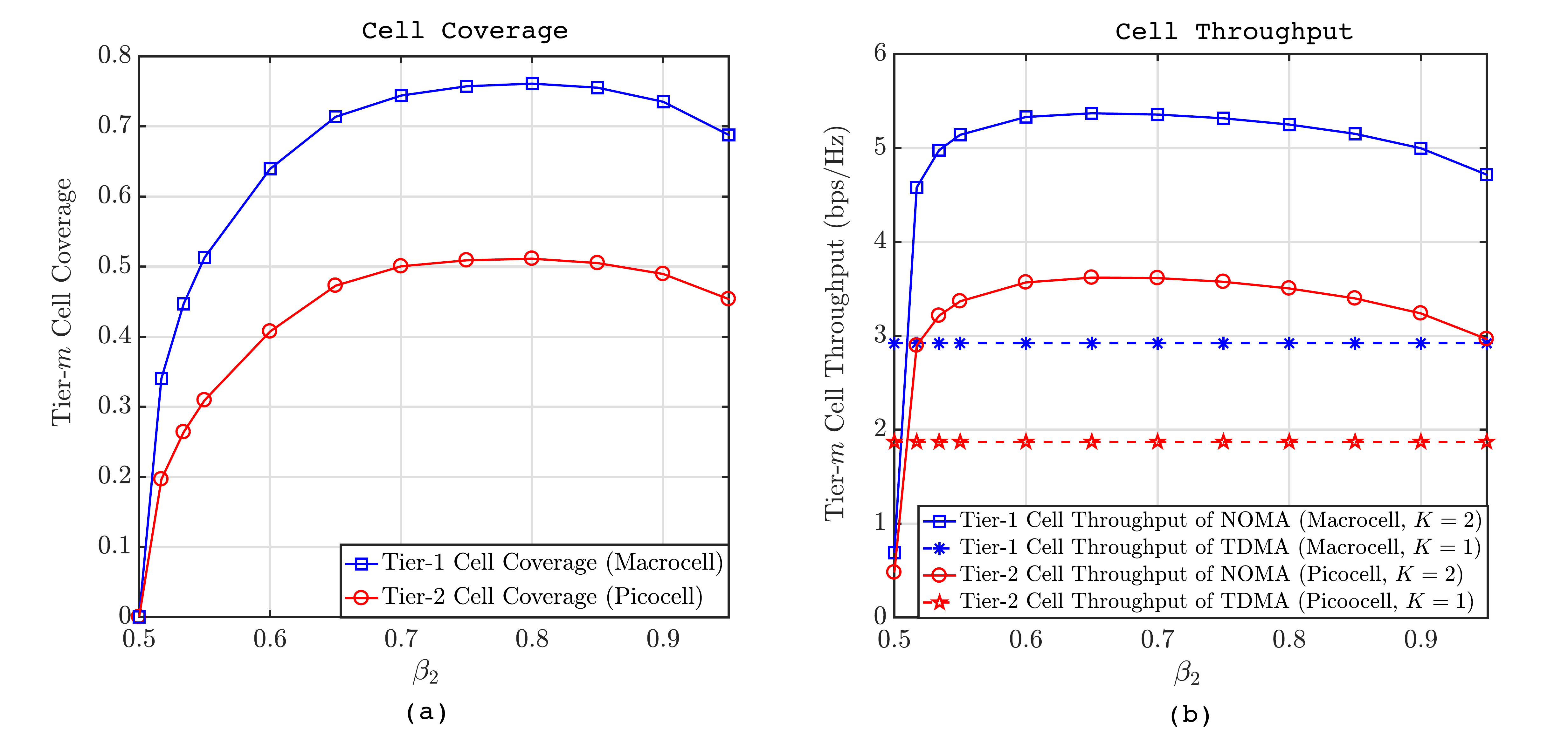}
	\caption{Numerical results of the tier-$m$ cell coverage and cell throughput for the non-coordinated NOMA scheme with $\mu=\lambda_2$ : (a) Tier-$m$ Cell Coverage $\frac{\rho_{m,1}+\rho_{m,2}}{2}$, (b) Tier-$m$ Cell Throughput $c_{m,1}+c_{m,2}$.}
	\label{Fig:CellCovThrNonCoopNOMA}
\end{figure}

In this subsection, we would like to show how the cell coverage and cell throughput of the BSs in each tier change with the power allocation schemes of the two users. According to the discussions in Section \ref{SubSec:CovAnaNonCooPNOMA}, for the non-coordinated NOMA scheme the constraint $\beta_1<\theta \beta_2$ must hold\footnote{This constraint is equivalent to the constraints $\beta_2>\frac{1}{1+\theta}$ and $\beta_1< \frac{\theta}{1+\theta}$ or $\beta_2>\frac{1}{2}$ and $\beta_{1}< \frac{1}{2}$ for $\theta=1$.} so that the two users do not fail to decode their signals merely due to the NOMA interference. This point can be verified by the simulation results in Fig. \ref{Fig:CellCovThrNonCoopNOMA} where the numerical results of the tier-$m$ cell coverage and cell throughput for the non-coordinated NOMA scheme are presented. As shown in Fig. \ref{Fig:CellCovThrNonCoopNOMA}, for $\beta_1=\beta_2=0.5$ the cell coverages are zero since the two users cannot decode their own signals just because of the NOMA interference from the other user. When $\beta_2$ starts to increase from 0.5, the tier-$m$ cell coverage initially increases, achieves to a maximum and then decreases. As indicated in Proposition \ref{Prop:OptimalPowerAllocationCoverage}, there exists an optimal power allocation vector $\mathbf{v}^{\star}_{\beta}=[\beta_1^{\star}\,\,\beta^{\star}_2]^{\mathrm{T}}$ that maximizes the tier-$m$ cell coverage. For example, we have $\mathbf{v}^{\star}_{\beta}=[0.2\,\, 0.8]^{\mathrm{T}}$ for the tier-$1$ cell coverage as shown in Fig. \ref{Fig:CellCovThrNonCoopNOMA} (a). 

\begin{figure}[t!]
	\centering
	\includegraphics[scale=0.35]{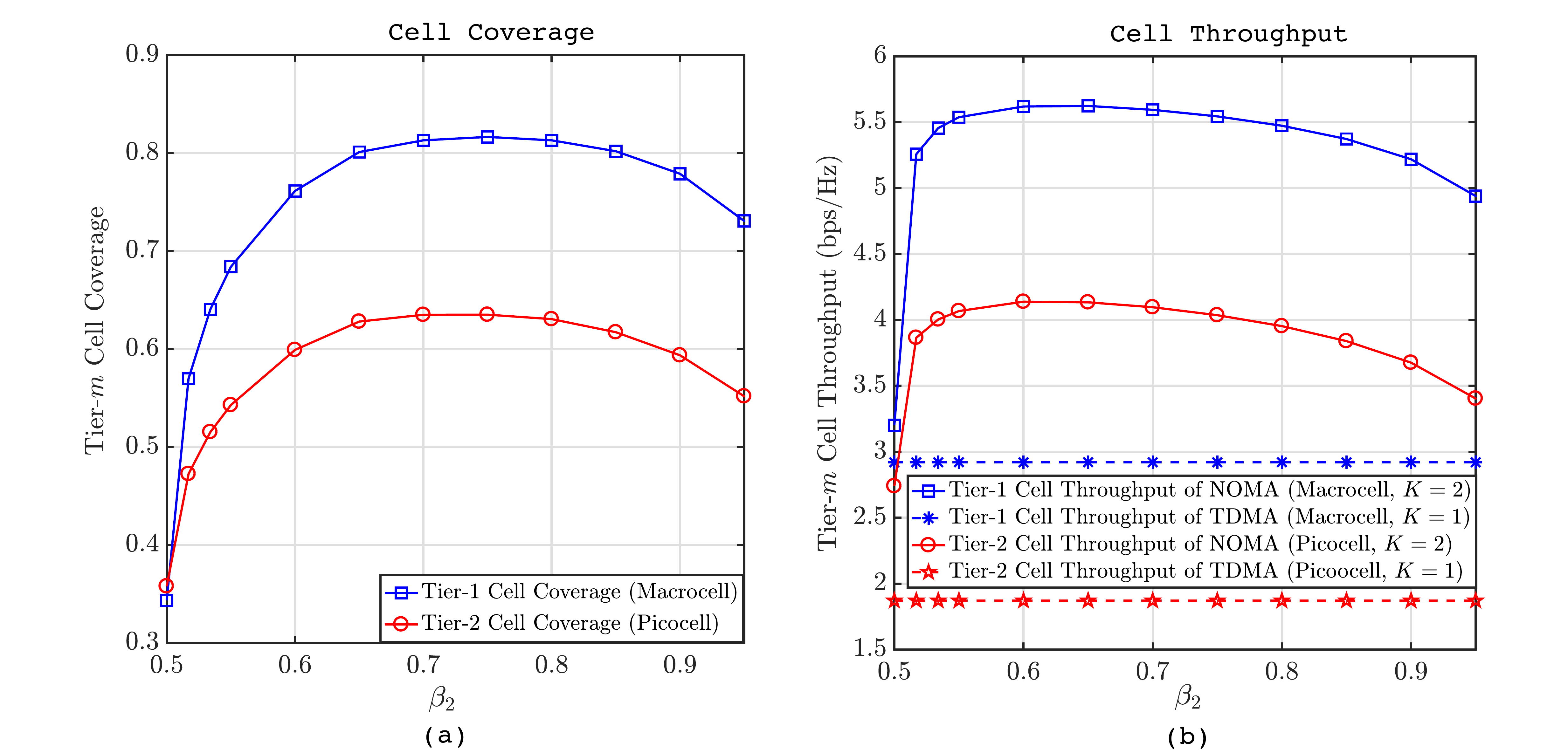}
	\caption{Numerical results of the tier-$m$ cell coverage and cell throughput for the coordinated JT-NOMA scheme with $\mu=\lambda_2$ : (a) Tier-$m$ cell coverage $\frac{\rho_{m,1}+\rho_{m,2}}{2}$, (b) Tier-$m$ cell throughput $c_{m,1}+c_{m,2}$.}
	\label{Fig:CellCovThrCoopNOMA}
\end{figure}

A similar phenomenon can also be observed in the case of the tier-$m$ cell throughput in Fig. \ref{Fig:CellCovThrNonCoopNOMA} (b), \ie the cell throughput for each tier can be largely reduced if $\beta_2\leq 0.5$ and there exists an optimal power allocation scheme that maximizes the tier-$m$ cell throughput, e.g., the tier-$2$ cell throughput maximizes when $\mathbf{v}^{\star}_{\beta}=[0.35\,\,0.65]^{\mathrm{T}}$. Most importantly, Fig. \ref{Fig:CellCovThrNonCoopNOMA} (b) indeed shows that NOMA outperforms TDMA for $0.5<\beta_2<1$, which validates our previous discussion that the sum of the link throughputs of the users is always strictly greater than the link throughput of the single user (OMA) scheme if the powers of the NOMA users are properly allocated. Consequently, NOMA is able to achieve a large throughput gain if the transmit powers are properly allocated among the users\footnote{Actually, we can show that optimization problems \eqref{Eqn:OptCovProb} and \eqref{Eqn:SumRateOptimizationNonCoopNOMA} are convex for $K=2$ so that their optimal solutions are unique.}.  In Fig. \ref{Fig:CellCovThrCoopNOMA}, we show the simulation results of the tier-$m$ coverage and cell throughput for the coordinated JT-NOMA scheme. As expected, the results in Fig.  \ref{Fig:CellCovThrCoopNOMA} are better than those in Fig. \ref{Fig:CellCovThrNonCoopNOMA} and they also can be maximized by optimizing the power allocation scheme between the two users.

\section{Conclusions}\label{Sec:Conclusions}
In the downlink transmission, a BS that performs the NOMA scheme to simultaneously serve multiple users can improve the SIR of the users if SIC is perfect. However, due to channel fading and inter-cell co-channel interference, SIC may fail at the user side so that NOMA may not provide a good SIR performance to all users at the same time. In this paper, the accurate explicit results of the coverage and link throughput of  the $K$ users associating with a BS in each tier for the non-coordinated NOMA scheme are firstly found and they show that non-coordinated NOMA can significantly degrade the coverage and the link throughput provided the transmit powers for the users are not allocated by following some fundamental constraints. In order to significantly improve the SIC and NOMA, we propose a coordinated JT-NOMA scheme in which all void BSs can do joint transmission to enhance the signal power of the farthest user in a cell. This coordinated JT-NOMA scheme is shown to significantly enhance the coverage and throughput performances of the users, especially in a dense network with a moderate user intensity. We finally show that the optimal power allocations for maximizing the tier-$m$ cell coverage and throughput indeed exist and they can be found by numerical techniques.  

\appendix[Proofs of Propositions]
\subsection{Proof of Proposition \ref{Prop:CCDF-DSIR-Tier-m}}\label{App:ProofPropCCDF-DSIR-Tier-m}
Due to the assumption of $H_{m,i,k}\sim\mathrm{Exp}(1)$, the CCDF of $\gamma_{m,k}$ in \eqref{Eqn:DSIR} for a given $x>0$ can be written as
\begin{align*}
F^{\mathsf{c}}_{\gamma_{m,k}}(x) &=\mathbb{E}\left[\exp\left(-x\frac{ I_{m,k}\|U_k\|^{\alpha}}{\beta_kP_m}\right)\right]=\mathbb{E}_{\|U_k\|^2}\left\{\mathbb{E}_{I_{m,k}}\left[\exp\left(-x\frac{ I_{m,k}\|U_k\|^{\alpha}}{\beta_kP_m}\right)\bigg|\|U_k\|^2 \right]\right\}\\
&=\mathbb{E}_{\|U_k\|^2}\left\{\prod_{l=1}^{M}\mathbb{E}\left[\exp\left(-\frac{x\omega_m\|U_k\|^{\alpha}}{\beta_k\omega_mP_m}\sum_{l,i:X_{l,i}\in\Phi_l\setminus X_{m,i}}\frac{\omega_l}{\omega_l}P_lV_{l,i}H_{l,i,k}\|X_{l,i}\|^{-\alpha}\right)\right]\right\}\\
&\stackrel{(a)}{=}\mathbb{E}_{\|U_k\|^2}\left\{\prod_{l=1}^{M}\mathbb{E}\left[\exp\left(-\frac{x\omega_m}{\beta_kP_m}\sum_{l,i:\widetilde{X}_{l,i}\in\widetilde{\Phi}_l\setminus\widetilde{X}_{m,i}}\frac{P_lV_{l,i}}{\omega_l}H_{l,i,k}\left(\frac{\|\widetilde{U}_k\|^{2}}{\|\widetilde{X}_{l,i}\|^{2}}\right)^{\frac{\alpha}{2}}\right)\right]\right\}\\
&\stackrel{(b)}{\gtrapprox} \mathbb{E}_{\|\widetilde{U}_k\|^2}\left\{\exp\left[-\pi\|\widetilde{U}_k\|^2\sum_{l=1}^{M}\nu_l\lambda_l\left(\frac{x \omega_mP_l}{\beta_k\omega_lP_m}\right)^{\frac{2}{\alpha}}\left(\frac{1}{\mathrm{sinc}(\frac{2}{\alpha})}-\int_{0}^{\left(\frac{\beta_k\omega_lP_m}{x\omega_m P_l}\right)^{\frac{\alpha}{2}}}\frac{\dif t}{1+t^{\frac{\alpha}{2}}}\right)\right]\right\}\\
&\stackrel{(c)}{=} \mathbb{E}_{\|\widetilde{U}_k\|^2}\left\{\exp\left[-\pi\widetilde{\lambda}_{\Sigma}\|\widetilde{U}_k\|^2\sum_{l=1}^{M}\nu_l\ell_{m,l}\left(\frac{x}{\beta_k}\right)\right]\right\},
\end{align*}
where $(a)$ follows from $\widetilde{X}_{l,i}\defn \omega^{-\frac{1}{\alpha}}_lX_{l,i}$, $\widetilde{U}_k\defn \omega^{-\frac{1}{\alpha}}_m U_k$, $\widetilde{\Phi}_l\defn\{\widetilde{X}_{l,i}\in\mathbb{R}^2: \widetilde{X}_{l,i}= \omega^{-\frac{1}{\alpha}}_lX_{l,i}, i\in\mathbb{N}_+\}$, and $\widetilde{\lambda}_{\Sigma}\defn\sum_{m=1}^{M}\omega^{\frac{2}{\alpha}}_m\lambda_m$, $(b)$ follows from the fact that the non-void BSs in the $l$th tier still can be accurately approximated by an independent PPP of intensity $\nu_l\lambda_l$ and this approximation leads to a tight lower bound result \cite{CHLLCW15,CHLLCW16}, $\|\widetilde{U}_k\|$ is the distance from $\widetilde{U}_k$ to $\widetilde{X}_{m,i}$ and the derived result in the proof of Theorem 4 in our previous work \cite{CHLKLF16} for the case of $H_{l,i,k}\sim\mathrm{Exp}(1)$, and $(c)$ is directly obtained by the definition of $\ell_{m,l}(\cdot)$ in \eqref{Eqn:ellfun-mk}. 

Since user $U_k$ adopts the BNBA scheme to associate with BS $X_{m,i}$ and it is the $k$th nearest user to $X_{m,i}$ among the $K$ scheduled users, the distribution of $\|\widetilde{U}_1\|^2$ can be equivalently written as $\|\widetilde{U}_1\|^2\stackrel{d}{=} D^{\min}_K$ where $\stackrel{d}{=}$ means the equivalence in distribution and $D^{\min}_K$ is the minimum RV among $K$ i.i.d. exponential RVs with parameter $\pi\widetilde{\lambda}_{\Sigma}$. Due to the memoryless property of exponential RVs, the distribution of $\|\widetilde{U}_2\|^2$ can be equivalently written as $\|\widetilde{U}_2\|^2\stackrel{d}{=}\|U_1\|^2+D^{\min}_{K-1}$ where $D^{\min}_{K-1}$ represents the minimum RV among $K-1$ i.i.d. RVs with parameter $\pi\widetilde{\lambda}_{\Sigma}$ \cite{CHLLCW16,MH12}, and note that $D^{\min}_{K}$ and $D^{\min}_{K-1}$ are independent. Accordingly,  $\|\widetilde{U}_k\|^2$ can be equivalently written as 
\begin{align*}
\|\widetilde{U}_k\|^2\stackrel{d}{=} \sum_{j=0}^{k-j} D^{\min}_{K-j},
\end{align*}
where all $D^{\min}_{K-j}$'s are independent and $D^{\min}_{K-j}\sim\mathrm{Exp}((K-j)\pi\widetilde{\lambda}_{\Sigma})$. Thus, for any $s>0$ we have
\begin{align*}
\mathbb{E}\left[e^{-s\|U_k\|^2}\right]=\prod_{j=0}^{k-j}\mathbb{E}\left[e^{-sD^{\min}_{K-j}}\right],\,\,\text{for }j\leq k,
\end{align*}
where $\mathbb{E}\left[e^{-sD^{\min}_{K-j}}\right]$ can be found as follows
\begin{align*}
\mathbb{E}\left[e^{-sD^{\min}_{K-j}}\right]&=\int_{0}^{\infty} e^{-sx}f_{D^{\min}_{K-j}}(s)\dif x=\int_{0}^{\infty} \pi(K-j)\widetilde{\lambda}_{\Sigma}e^{-[s+(K-j)\pi\widetilde{\lambda}_{\Sigma}]x}\\
&=\frac{(K-j)\pi\widetilde{\lambda}_{\Sigma}}{s+(K-j)\pi\widetilde{\lambda}_{\Sigma}}.
\end{align*}
Hence, we further have
\begin{align*}
\mathbb{E}\left[e^{-s\|U_k\|^2}\right]=\prod_{j=0}^{k-j}\left(\frac{(K-j)\pi\widetilde{\lambda}_{\Sigma}}{s+(K-j)\pi\widetilde{\lambda}_{\Sigma}}\right)
\end{align*}
Then $F^{\mathsf{c}}_{\gamma_{m,k}}(x)$ in \eqref{Eqn:CCDF-DSIR-mk} is acquired by substituting $s=\pi\widetilde{\lambda}_{\Sigma}\sum_{l=1}^{M}\nu_l\ell_{m,l}\left(\frac{x}{\beta_k}\right) $ into the result of $\mathbb{E}\left[e^{-s\|U_k\|^2}\right]$ found in above. 

\subsection{Proof of Proposition \ref{Prop:CovProbNOMA}}\label{App:ProfCovProbNOMA}
According to \eqref{Eqn:DefnCovProbUk}, we can rewrite $\rho_{m,k}$ as
\begin{align*}
\rho_{m,k} 
&=\mathbb{P}\left[\left(\beta_k-\theta\sum_{n=0}^{k-1}\beta_n\right)\gamma_{m,k}\geq \theta\beta_k,\cdots,\left(\beta_K-\theta\sum_{n=0}^{K-1}\beta_n\right)\gamma_{m,k}\geq \theta\beta_k\right]\\
&=\mathbb{P}\left[\gamma_{m,k}\geq \beta_k\max_{l\in\{k,\ldots,K\}}\left\{\frac{\theta}{\beta_l-\theta\sum_{n=0}^{l-1}\beta_n}\right\}\right]=\mathbb{P}\left[\gamma_{m,k}\geq \beta_k\vartheta_{k,K}\right],
\end{align*}
where the last equality follows from the condition $\beta_l>\theta\sum_{n=0}^{l-1}\beta_n$ for $l\in\{k,\dots,K\}$ and the definition of $\vartheta_{k,K}$ in \eqref{Eqn:EquivSIRThreshold-mk}. According to Proposition \ref{Prop:CCDF-DSIR-Tier-m}, $\rho_{m,k}$ can be written as
\begin{align*}
\rho_{m,k}=F^{\mathsf{c}}_{\gamma_{m,k}}\left(\beta_k\vartheta_{k,K}\right)
\end{align*}
and then substituting $\beta_k\vartheta_{k,K}$ into \eqref{Eqn:CCDF-DSIR-mk} leads to the tight lower bound in \eqref{Eqn:CovProbUserk}. Also, as $\mu\rightarrow\infty$, all $\nu_l$'s converge to one so that all BSs are not void and thus the location correlations among the non-void BSs no longer exist. Therefore, $\lim_{\mu\rightarrow\infty}\rho_{m,k}$ is equal to \eqref{Eqn:LowLimitCovProbUserk} that is the lower bound in \eqref{Eqn:CovProbUserk} with $\nu_l=1$ for all $l\in\mathcal{M}$ according to the proof of Proposition \ref{Prop:CCDF-DSIR-Tier-m}.

\subsection{Proof of Proposition \ref{Prop:LinkThroughputNonCoopNOMA}}\label{App:ProofLinkThroughputNonCoopNOMA}
First note that for any $x,y,z>0$ we have the following identity:
\begin{align*}
\log\left(1+\frac{x}{y+z}\right) = \log\left(1+\frac{x+y}{z}\right)-\log\left(1+\frac{y}{z}\right).
\end{align*}
Accordingly, $c_{m,k}$ in \eqref{Eqn:DefnLinkThrough-mk} can be rewritten as
\begin{align}\label{Eqn:ProofLinkThrputUserk}
c_{m,k}=\mathbb{E}\left[\log\left(1+\frac{\sum_{n=1}^{k}\beta_n}{\beta_k}\gamma_{m,k}\right)-\log\left(1+\frac{\sum_{n=0}^{k-1}\beta_n}{\beta_k}\gamma_{m,k}\right)\bigg| \gamma_{m,k}\geq \beta_k\vartheta_{k+1,K}\right],
\end{align}
and for $a,b>0$ we have the following
\begin{align*}
&\mathbb{E}\left[\log(1+a\gamma_{m,k})|\gamma_{m,k}\geq b\right]=\int_{0}^{\infty}\mathbb{P}\left[\log(1+a\gamma_{m,k})\geq x|\gamma_{m,k}\geq b\right]\dif x\\
&=\int_{0}^{\infty} \frac{\mathbb{P}\left[\gamma_{m,k}\geq \frac{y}{a},\gamma_{m,k}\geq b\right]}{\mathbb{P}[\gamma_{m,k}\geq b]}\frac{\dif y}{1+y}=\log\left(1+ab\right)+\int_{ab}^{\infty}\frac{F^{\mathsf{c}}_{\gamma_{m,k}}(y/a)}{F^{\mathsf{c}}_{\gamma_{m,k}}(b)}\frac{\dif y}{1+y}.
\end{align*}
Then using Proposition \ref{Prop:CCDF-DSIR-Tier-m} and letting $a=\frac{\sum_{n=1}^{k}\beta_n}{\beta_k}$ and $b=\beta_k\vartheta_{k+1,K}$ lead to the following results of $c_{m,k}$:
\begin{align*}
c_{m,k}=&\log\left(1+\vartheta_{k+1,K}\sum_{n=1}^{k}\beta_n\right)+\int_{\vartheta_{k+1,K}\sum_{n=1}^{k}\beta_n }^{\infty}\frac{F^{\mathsf{c}}_{\gamma_{m,k}}(y\beta_k/\sum_{n=1}^{k}\beta_n)\dif y}{F^{\mathsf{c}}_{\gamma_{m,k}}(\beta_k\vartheta_{k+1,K})(1+y)}\\
&-\log\left(1+\vartheta_{k+1,K}\sum_{n=0}^{k-1}\beta_n\right)-\int_{\vartheta_{k+1,K}\sum_{n=0}^{k-1}\beta_n }^{\infty}\frac{F^{\mathsf{c}}_{\gamma_{m,k}}(y\beta_k/\sum_{n=0}^{k-1}\beta_n)\dif y}{F^{\mathsf{c}}_{\gamma_{m,k}}(\beta_k\vartheta_{k+1,K})(1+y)}\\
=&\log\left(1+\frac{\beta_k\vartheta_{k+1,K}}{\vartheta_{k+1,K}\sum_{n=0}^{k-1}\beta_n+1}\right)+\int_{\beta_k\vartheta_{k+1,K}}^{\infty}\left[\frac{\eta^2_kF^{\mathsf{c}}_{\gamma_{m,k}}(z)/F^{\mathsf{c}}_{\gamma_{m,k}}(\beta_k\vartheta_{k+1,K})}{(\beta_k+z\sum_{n=1}^{k}\beta_n)(\beta_k+z\sum_{n=0}^{k-1}\beta_n)}\right]\dif z.
\end{align*} 
Thus, the tight lower bound in \eqref{Eqn:LinkThrouhgput-mk} can be readily acquired by using the bound in \eqref{Eqn:CCDF-DSIR-mk} to find the results of $F^{\mathsf{c}}_{\gamma_{m,k}}(\beta_k\vartheta_{k+1,K})$ and $F^{\mathsf{c}}_{\gamma_{m,k}}(z)$.  Now consider $k=K$ and $c_{m,K}$ can be found as
\begin{align}
c_{m,K}&=\mathbb{E}\left[\log\left(1+\frac{\sum_{n=1}^{K}\beta_n}{\beta_K}\gamma_{m,K}\right)-\log\left(1+\frac{\sum_{n=0}^{K-1}\beta_n}{\beta_K}\gamma_{m,K}\right)\right]\nonumber\\
&=\mathbb{E}\left[\log\left(1+\frac{\gamma_{m,K}}{\beta_K}\right)\right]-\mathbb{E}\left[\log\left(1+\frac{(1-\beta_K)}{\beta_K}\gamma_{m,K}\right)\right]\label{Eqn:ProofLinkThrputUserK2}\\
&=\int_{0}^{\infty} \frac{F^{\mathsf{c}}_{\gamma_{m,K}}(y\beta_K)-F^{\mathsf{c}}_{\gamma_{m,K}}(y\beta_K/(1-\beta_K))}{(1+y)}\dif y.\nonumber
\end{align}
Then the tight lower bound in \eqref{Eqn:LinkThroughput-mK} can be readily obtained by substituting the tight lower bounds on $F^{\mathsf{c}}_{\gamma_{m,K}}(x)$ in \eqref{Eqn:CCDF-DSIR-mk} with $x=y/\beta_K$ and $x=y\beta_K/(1-\beta_K)$ into the expression of $c_{m,K}$ in above.

\subsection{Proof of Proposition \ref{Prop:CovProbCooPNOMA}}\label{App:ProfCovProbCooPNOMA}
According to the proof of Proposition \ref{Prop:CovProbNOMA}, $\rho_{m,k}$ for $k\in\{1,2,\ldots,K-1\}$ defined in \eqref{Eqn:DefnCovProbCoopNOMA}, and assuming all void BSs in the $l$th tier to be an independent PPP of intensity $\nu_l\lambda_l$ we have
\begin{align*}
\rho_{m,k} &=\mathbb{P}\left[\left(\beta_k-\theta\sum_{n=0}^{k-1}\beta_n\right)\gamma_{m,k}\geq \theta\beta_k,\ldots,\left(\beta_K-\theta\sum_{n=0}^{K-1}\beta_n\right)\gamma_{m,k}\geq \theta\beta_k-\frac{S_{m,k}}{I_{m,k}}\right]\\
&\stackrel{(a)}{=}\mathbb{P}\left[\gamma_{m,k}\geq \max_{l\in\{k,\ldots,K\}}\left\{\frac{\beta_k}{(\beta_l-\theta\sum_{n=0}^{l-1}\beta_n)}\left[\theta-\frac{S_{m,k}}{\beta_kI_{m,k}}\mathds{1}(l=K)\right]\right\}\right]\\
&\stackrel{(b)}{=}\mathbb{P}\left[\gamma_{m,k}\geq \max_{l\in\{k,\ldots,K-1\}}\left\{\frac{\beta_k\theta}{(\beta_l-\theta\sum_{n=0}^{l-1}\beta_n)}\right\}\right]=F^{\mathsf{c}}_{\gamma_{m,k}}\left(\beta_k\vartheta_{k,K-1}\right)\\
&\stackrel{(c)}{\gtrapprox}  \prod_{j=0}^{k-1} \frac{(K-j)}{(K-j)+\sum_{l=1}^{M}\nu_l\ell_{m,l}\left(\vartheta_{k,K-1}\right)},
\end{align*}
where $(a)$ is due to the constraint $\theta\sum_{n=0}^{l-1}\beta_n< \beta_l< 1$ for $l\in\{k,\ldots,K\}$, $(b)$ is due to the constraint that $\beta_l-\theta\sum_{n=0}^{l-1}\beta_n< \beta_K-\theta\sum^{K-1}_{n=0}\beta_n$ (\ie $\beta_l< \beta_K-\theta\sum^{K-1}_{n=l}\beta_n$), and $(c)$ follows from the result in \eqref{Eqn:CovProbUserk}. Hence, the tight lower bound in \eqref{Eqn:CoverProbCoopNOMAUserk} is obtained. 

For the $K$th user, its coverage probability can be expressed as
\begin{align*}
\rho_{m,K}&=\mathbb{P}\left[\frac{\beta_KP_mH_{m,i,K}\|U_K\|^{-\alpha}+S_{m,K}}{(\sum_{l=0}^{K-1}\beta_l)P_mH_{m,i,K}\|U_K\|^{-\alpha}+I_{m,K}}\geq \theta\right]=\mathbb{P}\left[H_{m,i,K}\geq \frac{\|U_{K}\|^{\alpha}\left(\theta I_{m,K}-S_{m,K}\right)}{P_m\left(\beta_K-\theta\sum_{l=0}^{K-1}\beta_l\right)}\right]\\
&=\mathbb{E}\left[\exp\left(-\frac{\|U_{K}\|^{\alpha}\vartheta_{K,K}}{P_m}\left( I_{m,K}-\frac{S_{m,K}}{\theta}\right)\right)\right]\\
&=\mathbb{E}\left[\exp\left(-\frac{\omega_m\|U_{K}\|^{\alpha}\vartheta_{K,K}}{\omega_mP_m}\sum_{l,j:X_{l,j}\in\Phi\setminus X_{m,i}}\frac{V'_{l,j}\omega_lP_lH_{l,j,K}}{\omega_l\|X_{l,j}-U_K\|^{\alpha}}  \right)\right],
\end{align*}
where $V'_{l,j}\defn V_{l,j}\left(1+\frac{1}{\theta}\right)-\frac{1}{\theta}$.  Since location correlations among the non-void and void BSs induced by user association are fairly weak \cite{CHLLCW15}, we can find the approximated $\rho_{m,K}$ by assuming all $V'_{l,j}$'s are independent so that we can have the following approximation:
\begin{align*}
\rho_{m,K} \stackrel{(c)}{\approx}& \mathbb{E}\left[\exp\left(-\pi\|\tilde{U}_{K}\|^2\sum_{l=1}^M \lambda_l \int_{1}^{\infty}\mathbb{E}\left[1-e^{-\frac{\vartheta_{K,K} \omega_mP_lV'_{l,j}H_{l,j,K}}{\omega_lP_mr^{\frac{\alpha}{2}}}}\right]\dif r\right)\right] \\
=& \mathbb{E}\bigg[\exp\bigg(-\pi\|\tilde{U}_{K}\|^2\sum_{l=1}^M \left(\frac{\omega_mP_l}{\omega_lP_m}\right)^{\frac{2}{\alpha}} \lambda_l \int_{(\frac{\omega_lP_m}{\omega_mP_l})^{\frac{2}{\alpha}}}^{\infty}\bigg\{\nu_l\mathbb{E}\left[1-e^{-\frac{\vartheta_{K,K}H}{x^{\frac{\alpha}{2}}}}\right]\\
&+(1-\nu_l)\mathbb{E}\left[1-e^{\frac{\vartheta_{K,K} H}{\theta x^{\frac{\alpha}{2}}}}\right]\bigg\}\dif x\bigg)\bigg]
\end{align*}
\begin{align*}
\stackrel{(d)}{=}&\mathbb{E}_{\|\widetilde{U}_K\|^2}\left\{\exp\left(-\pi\widetilde{\lambda}_{\Sigma}\|\widetilde{U}_K\|^2\left[\sum_{l=1}^{M}\nu_l\ell_{m,l}\left(\vartheta_{K,K}\right)+(1-\nu_l)\widetilde{\ell}_{m,l}\left(\frac{\vartheta_{K,K}}{\theta}\right)\right]^+\right)\right\},
\end{align*}
where $(c)$ follows the proof technique introduced in the proof of Proposition \ref{Prop:CCDF-DSIR-Tier-m} and $(d)$ is obtained by the definitions of $\ell_{m,l}(\cdot)$ and $\widetilde{\ell}_{m,l}(\cdot)$. Then the result in \eqref{Eqn:CoverProbCoopNOMAUserK} can be obtained by applying the results in the proof of Proposition \ref{Prop:CCDF-DSIR-Tier-m}. Finally, \eqref{Eqn:CoverProbCoopNOMAUserK} reduces to \eqref{Eqn:CoverProbCoopNOMAUserkNoVoid} since all $\nu_l$'s converge to 1 as $\mu$ goes to infinity.

\subsection{Proof of Proposition \ref{Prop:LinkThoughput_CoopNOMA}}\label{App:ProofLinkThroughputCoopNOMA}
For $\beta_l\in(\theta\sum_{n=0}^{l-1}\beta_n,\beta_K-\theta\sum_{n=0}^{K-1}\beta_n)$ with $\beta_K\in(\theta\sum_{n=0}^{K-1}\beta_n,1)$, using the result of Appendix \ref{App:ProofLinkThroughputNonCoopNOMA} we can rewrite $c_{m,k}$ in \eqref{Eqn:DefnLinkThroughmkCoopNOMA} for $k\in\{1,2,\ldots,K-2\}$ as follows
\begin{align*}
c_{m,k} \defn\mathbb{E}\left[\log\left(1+\frac{\sum_{n=1}^{k}\beta_n}{\beta_k}\gamma_{m,k}\right)-\log\left(1+\frac{\sum_{n=0}^{k-1}\beta_n}{\beta_k}\gamma_{m,k}\right)\bigg| \gamma_{m,k}\geq\beta_k\vartheta_{k+1,K-1}\right],
\end{align*}
which is similar to $c_{m,k}$ found in \eqref{Eqn:ProofLinkThrputUserk}. For the $(K-1)$th user, its link throughput can be written as
\begin{align*}
c_{m,K-1} 
&=\mathbb{E}\left[\log\left(1+\frac{\sum_{n=0}^{K-1}\beta_n}{\beta_{K-1}}\gamma_{m,K-1}\right)-\log\left(1+\frac{\sum_{n=0}^{K-2}\beta_n}{\beta_{K-1}}\gamma_{m,K-1}\right)\right]
\end{align*}
for $\beta_l\in[\theta\sum_{n=0}^{l-1}\beta_n,\beta_K-\theta\sum_{n=0}^{K-1}\beta_n]$ with $\beta_K\in[\theta\sum_{n=0}^{K-1}\beta_n,1]$ and it is similar to $c_{m,K}$ defined in \eqref{Eqn:DefnLinkThrough-mK}. Hence, the result in \eqref{Eqn:LinkThrough-mkCoopNOMA1} can be readily obtained from \eqref{Eqn:LinkThrouhgput-mk} by changing $K$ to $K-1$, and the result in \eqref{Eqn:LinkThrough-mkCoopNOMA2} can be found directly from the result in \eqref{Eqn:LinkThroughput-mK} by replacing $K$ with $K-1$. Finally, the link throughput of the $K$th user can be expressed as
\begin{align*}
c_{m,K}&=\mathbb{E}\left[\log\left(1+\frac{\gamma_{m,K}+\frac{S_{m,K}}{I_{m,K}}}{(\sum_{n=0}^{K-1}\beta_n)\frac{\gamma_{m,K}}{\beta_K}+1}\right)\right]=\int_0^{\infty}\mathbb{P}\left[\frac{\gamma_{m,K}+\frac{S_{m,K}}{I_{m,K}}}{(\sum_{n=0}^{K-1}\beta_n)\frac{\gamma_{m,K}}{\beta_K}+1}\geq \theta\right]\frac{\dif \theta}{1+\theta}\\
&=\int_{0}^{\frac{\beta_K}{\sum_{n=0}^{K-1}\beta_n}}\mathbb{P}\left[\gamma_{m,K}\geq \frac{\theta-\frac{S_{m,K}}{I_{m,K}}}{1-(\sum_{n=0}^{K-1}\beta_n)\frac{\theta}{\beta_K}}\right]\dif \theta=\int_{0}^{\frac{\beta_K}{\sum_{n=0}^{K-1}\beta_n}} \frac{\rho_{m,K}(\theta)}{1+\theta}\dif \theta,
\end{align*}
where $\rho_{m,K}(\theta)$ is the coverage probability of the $K$th user already given in \eqref{Eqn:CoverProbCoopNOMAUserK}. Thus, substituting \eqref{Eqn:CoverProbCoopNOMAUserK} into $c_{m,K}$ above yields the approximated result in \eqref{Eqn:LinkThrough-mkCoopNOMA3}.  

\subsection{Proof of Proposition \ref{Prop:OptimalPowerAllocationCoverage}}\label{App:ProofOptimalPowerAllocationCoverage}
According to $\vartheta_{k,K}$ defined in \eqref{Eqn:EquivSIRThreshold-mk}, we readily know $\vartheta_{k+1,K}\leq \vartheta_{k,K} \leq \vartheta_{k-1,K}$ and this follows 
$\rho_{m,k}\left(\vartheta_{k-1,K}\right)\leq \rho_{m,k}\left(\vartheta_{k,K}\right) \leq   \rho_{m,k}\left(\vartheta_{k+1,K}\right)$
since $\ell_{m,l}(x)$ is a monotonically increasing function of $x$ as shown in \eqref{Eqn:ellfun-mk} and thus $\rho_{m,k}(x)$ is monotonically decreasing along $x$. 
In other words, we must have $\rho_{m,k}\left(\vartheta_{1,K}\right)\leq\cdots\leq \rho_{m,k}\left(\vartheta_{k,K}\right)\leq\cdots \leq \rho_{m,k}\left(\vartheta_{K,K}\right)$ and this follows that  $\sum_{k=1}^{K}\rho_{m,k}\left(\vartheta_{k,K}\right)\leq \sum_{k=1}^{K}\rho_{m,k}\left(\vartheta_{K,K}\right)$. Thus, $\frac{1}{K}\sum_{k=1}^{K}\rho_{m,k}$ is continuous and bounded for all $\mathbf{v}_{\beta}\in(0,1)^K$ because $\frac{1}{K}\sum_{k=1}^{K}\rho_{m,k}\left(\vartheta_{K,K}\right)$ is bounded for any $\vartheta_{K,K}$ that is determined by $\theta$ and $\mathbf{v}_{\beta}$. In other words, $\frac{1}{K}\sum_{k=1}^{K}\rho_{m,k}$ is also continuous and bounded for any $\mathbf{v}_{\beta}\in\mathcal{V}_{\beta}(\theta)\subset(0,1)^K$. Also, note that $\mathcal{V}_{\beta}(\theta)$ in \eqref{Eqn:PowerAllSetNonCoopNOMA} and $\mathcal{V}_{\beta}(\theta)$ in \eqref{Eqn:PowerAllSetCoopNOMA} are both a polyhedron so that they are compact. Accordingly, there must exist an optimal vector $\mathbf{v}_{\beta}^{\star}\in\mathcal{V}_{\beta}(\theta)$ that maximizes $\frac{1}{K}\sum_{k=1}^{K} \rho_{m,k}$ based on the Weierstrass theorem \cite{DPB16}. 


\bibliographystyle{ieeetran}
\bibliography{IEEEabrv,Ref_CoopNOMA}

\end{document}